\documentclass[aps,prb,twocolumn,showpacs,superscriptaddress,shopacs]{revtex4}

\usepackage{graphicx}
\usepackage{bm}
\usepackage{gensymb}
\usepackage[none]{hyphenat}
\usepackage{epstopdf}
\usepackage{multirow}
\usepackage{float}
\usepackage{tabularx}
\usepackage{gensymb}
\usepackage{color}
\usepackage{siunitx}

\begin{document}

\title{Modulated magnetism and anomalous electronic transport in Ce$_3$Cu$_4$As$_4$O$_2$}

\author{Jiakui K. Wang $^{\footnote[1]{These authors contributed equally to the project.}}$}
\affiliation{Department of Physics and Astronomy, Rice University, Houston, TX 77005, USA}

\author{Shan Wu $^*$ }
\affiliation{Department of Physics and Astronomy and Institute for Quantum Matter, Johns Hopkins University, Baltimore, MD 21218, USA }

\author{Yiming Qiu}
\affiliation{NIST Center for Neutron Research, National Institute of Standards and Technology, Gaithersburg, MD 20899, USA}
\affiliation{Department of Materials Science, University of Maryland, College Park, MD 20742, USA}

\author{Jose A. Rodriguez-Rivera}
\affiliation{NIST Center for Neutron Research, National Institute of Standards and Technology, Gaithersburg, MD 20899, USA}
\affiliation{Department of Materials Science, University of Maryland, College Park, MD 20742, USA}

\author{Qingzhen Huang}
\affiliation{NIST Center for Neutron Research, National Institute of Standards and Technology, Gaithersburg, MD 20899, USA}
\affiliation{Department of Materials Science, University of Maryland, College Park, MD 20742, USA}

\author{C. Broholm}
\affiliation{Department of Physics and Astronomy and Institute for Quantum Matter, Johns Hopkins University, Baltimore, MD 21218, USA }
\affiliation{Department of Materials Science and Engineering, Johns Hopkins University, Baltimore, MD 21218, USA}

\author{E. Morosan}
\affiliation{Department of Physics and Astronomy, Rice University, Houston, TX 77005, USA}

\date{\today}

\begin{abstract}

The complex magnetism and transport properties of tetragonal Ce$_3$Cu$_4$As$_4$O$_2$ were examined through neutron scattering and physical properties measurements on polycrystalline samples. The lamellar structure consists of alternating layers of $\rm CeCu_4As_4$ with a single square Ce lattice and oxygen-linked Ce bi-layer $\rm Ce_2O_2$. Extending along $\bf c$, a tube-like Fermi surface from DFT calculations points to a quasi-two-dimensional electronic system. Peaks in the specific heat at the Ne\'{e}l temperature $T_{N}=24$~K, $T_{2}~=~16$~K and $T_{3}~=~1.9$~K indicate three magnetic phase transitions or distinct cross-over phenomena. For $T<T_{N}$ neutron diffraction indicates the development of ferromagnetic \textit{ab} sheets for both Ce sites, with alternating polarization along $\bf{c}$, a wave vector ${\bf k}_{1}={\bf c}^*$. For $T<T_{2}$, quasi-two-dimensional low-energy spin fluctuations with ${\bf k}_{2}=\frac{1}{2}{\bf a}^*$ and polarized perpendicular to ${\bf k}_{2}$ are suppressed. The data are consistent with quasi-two-dimensional antiferromagnetic order in the $\rm CeCu_4As_4$ planes polarized along the ${\bf k}_{2}$  wave vector. $T_{3}$ marks a spin-flop transition where the ${\bf k}_{1}$ staggered magnetization switches to in-plane polarization. While the narrow 4f bands lie deep below the Fermi surface, there are significant transport anomalies associated with the transitions; in particular a substantial reduction in resistivity for $T<T_{N}$. At $T=100$~mK the ${\bf k}_1$ modulated staggered moment  is $0.85~\mu_B$, which matches the $0.8~\mu_B$ saturation magnetization achieved for H $~=~7$~T at $T~=~2$~K. From low T Lorentzian fits  the correlation length is in excess of 75 \AA. We argue the unusual sequence of magnetic transitions results from competing interactions and anisotropies for the two Ce sites.
\end{abstract}


\maketitle


\section{Introduction}

As spin degeneracy is lifted through crystal field and exchange interactions, intermetallic compounds containing rare earth ions display intricate thermodynamic, magnetic, and transport anomalies. In the Kondo effect dilute rare earth impurities in a metal give rise to a minimum in the resistivity due to resonant electron scattering from the impurity spin\cite{kondo}$^,$ \cite{coqblin}.  When the rare-earth ions form a full crystalline lattice and when hybridization with more dispersive bands is sufficiently strong, a full band gap can open in so-called Kondo insulators which may support topologically protected surface states\cite{Fisk}$^,$\cite{Dzero}$^,$\cite{Mason}. The Kondo lattice regime between these limits is defined by an intricate balance between inter-site super-exchange and intra-site Kondo screening\cite{Doniach}. When the former interactions prevail, magnetic order occurs concomitant with transport anomalies, while the ground state in the latter case is a heavy Fermi Liquid (FL). At the quantum critical point (QCP) where these phases meet, there are non-Fermi-Liquid (NFL) characteristics and in some cases superconductivity\cite{Stewart}$^,$\cite{Coleman}.

The search for materials to expose this strongly correlated regime recently led to the discovery of a family of rare earth bearing quasi-two-dimensional metals of the form $\rm R_3T_4As_4O_{2-\delta}$ \cite{kaiser, cava, jiakui} that is structurally related to the iron superconductors. Here R indicates a rare earth ion and T a transition metal ion. While T appears to be non-magnetic in this structure,\cite{jiakui} it strongly influences the sequence of transitions so there are three transitions for $\rm Ce_3Cu_4As_4O_2$ but just a single low temperature transition for $\rm Ce_3Ni_4As_4O_2$. The presence of {\em two} distinct rare earth sites appears to underlie the complicated magnetic and transport properties of these materials. Focusing on $\rm Ce_3Cu_4As_4O_2$, we seek in this paper to understand the physics underlying the previously reported sequence of phase transitions in this class of materials. While the absence of single crystalline samples limits the specificity of our conclusions, the combination of thermodynamic, transport and neutron scattering data that we shall report, as well as ab-initio band structure calculations, provides a first  atomic scale view of the intricate electronic properties of these materials. It is also apparent however, that the complexity of the material is such that a full understanding will require much more detailed experiment that can only be carried out on single crystalline samples.

So far our experiments indicate that incompatible in-plane magnetic interactions and magnetic anisotropies produce separate as well as coordinated magnetic phase transitions reminiscent of phase transitions in magnetic multilayer thin-film structures. The strongly anisotropic nature of the cerium spins and/or their interactions also plays an important role by allowing for a thermodynamic phase transition for an isolated 2D layer. The highest temperature transition at $T_{N}=24$~K is to an antiferromagnetic (AFM) stacking of ferromagnetically (FM) aligned tetragonal layers of spins oriented along the c axis, with characteristic magnetic wave vector ${\bf k}_1=1.0(2){\bf c^*}$. This order is predominantly associated with the $\rm Ce_2O_2$ layers. The $T_2$ = 16 K transition is associated with the loss of low energy spin fluctuations in the square sublattice $\rm CeCu_4As_4$. While we do not have direct diffraction evidence for this, the paramagnetic spin fluctuations indicate that the spin structure in $\rm CeCu_4As_4$ layers for $T<T_{2}$ has a characteristic wave vector ${\bf k}_2=0.50(2){\bf a}^*$ with spins along ${\bf k}_2$. Such longitudinally striped structures are also found among iron spins in the parent compounds of iron superconductors\cite{su}$^,$\cite{Clarina}, and is supported by our DFT calculations. The lowest temperature  transition at $T_{3}=1.9$~K is associated with the development of a component of the ${\bf k}_1$-type magnetic order polarized within the basal plane. Indeed the data is consistent with the rotation of the entire polarization of $\rm Ce_2O_2$ layers into the basal plane and perpendicular to an anisotropic striped AFM of the $\rm CeCu_4As_4$ layers. The paper thus exposes an intricate interplay between distinct forms of rare earth magnetism in the two different layers that make up  $\rm Ce_3Cu_4As_4O_2$.

After the methods section, our experimental results and initial observations are presented in section~\ref{results} followed by analysis in section~\ref{analysis}. Section~\ref{discussion} draws together a physical picture of $\rm Ce_3Cu_4As_4O_2$ and in the concluding section (\ref{conclusion}) we put the results into the broader context of the $\rm R_3T_4As_4O_2$ family of compounds and rare earth based strongly correlated electron systems in general.

\section{Methods}

\label{methods}

\subsection{Synthesis and bulk measurements}

Polycrystalline Ce$_3$Cu$_4$As$_4$O$_2$ was synthesized using a previously described solid state method\cite{jiakui}. The sample employed for neutron scattering was a loose powder with a total mass of 5.8 g.

Magnetization measurements for temperatures between 2 K and 300 K were performed in a Quantum Design (QD) Magnetic Property Measurement System (MPMS) \cite{quantumdesign} in magnetic fields up to 7 T. Specific heat measurements were performed in a QD Physical Property Measurement System (PPMS) \cite{quantumdesign}, using an adiabatic relaxation method for temperature down to 0.4 K and fields up to 9 T. DC electrical resistivity was measured on dense samples in the QD PPMS \cite{quantumdesign}, using a standard four point contact method.

\subsection{Neutron scattering}
All neutron scattering experiments were conducted on instrumentation at the NIST Center for Neutron Research. For determination of the chemical structure powder diffraction data was acquired on BT1 using the Ge(311) monochromator  $\lambda = 1.5398 ~ \AA$, with 60$^\prime$ in-pile collimation and the standard 20$^\prime$ and 7$^\prime$ collimation before and after the sample respectively. In this measurement the powder sample was held in a thin walled vanadium can with $^4$He as exchange gas and the sample was cooled by a closed cycle compressor system.
Rietveld refinement of the BT1 data was carried out using the General Structure Analysis System (\textit{GSAS}) \cite{gsas}.

To detect magnetic diffraction from the relatively small moment magnetism of Ce$_3$Cu$_4$As$_4$O$_2$, high intensity diffraction data were acquired for temperatures between 50 mK and 45 K on the Multi-Axis Crystal Spectrometer (MACS) at NIST\cite{macs}. For these long wave length measurements the powder was held in a thin-walled aluminum can with $^4$He as exchange gas. A total of three different MACS experiments employing as many cryogenic systems contributed to the paper: In May 2013 we used an "Orange" $^4$He flow cryostat with access to temperatures from 1.5 K to 50 K. In November 2013 we used an Oxford Instruments dilution fridge with an 11.5 T magnet for the temperature range from 0.05 K to 14 K. Finally in May 2014 a dilution insert with access to temperatures from 0.1 K to 40 K was employed.

The incident and final neutron energies on MACS were defined by PG(002) monochromatization to be 5 meV. The full double focusing configuration of the MACS monochromator was used for temperature  scans while higher resolution data were collected with the monochromator set for vertical focusing only and the beam width limited to 10 cm at the pre-monochromator beam aperture. The horizontal angular divergence of the incident neutron beam was $3.8^\circ$ and  $0.6^\circ$ for the doubly and singly focused configurations respectively.

A second set of energy integrating detectors on MACS were utilized in this experiment. These ``two-axis'' detectors are behind the PG(002) double bounce analyzers and detect scattered neutrons irrespectively of the final energy of scattering. With an incident energy $E_i= 5$~meV and a beryllium filter between the powder sample and the analyzer the measured quantity can be written as follows\cite{Squires}:
\begin{equation}
\frac{d\sigma}{d\Omega}=r_0^2\int_0^{E_i}\sqrt{1-\frac{\hbar\omega}{E_i}}|\frac{g}{2}F(Q_\omega)|^2 2\tilde{\cal S}(Q_\omega,\omega) \hbar d\omega
\label{energyintegrated}
\end{equation}
Here $Q_\omega=|{\bf k}_i-{\bf k}_f|$, ${\bf k}_i$ and ${\bf k}_f$ being the wave vectors of incident and scattered neutrons respectively, and $\tilde{\cal S}(Q_\omega,\omega)$ is the spherically averaged dynamic correlation function defined so as to include elastic scattering.

For the ultimate sensitivity to weak temperature-dependent scattering we plot and analyze temperature difference intensity data from the spectroscopic detectors on MACS. While this removes $T-$independent scattering on average, thermal expansion produces peak derivative line shapes in place of nuclear Bragg peaks. These features can be accounted for quantitatively based on thermal expansion coefficients and a Rietveld refinement \cite{rietveld} of the nuclear diffraction data as follows.

Using angular variables as is customary for monochromatic beam diffraction data, the Rietveld refined nuclear diffraction pattern is written as follows:
\begin{equation}
{\cal I}(\theta)={\cal C}\sum_{\tau}\frac{I(\tau)}{2\sigma(\tau)}g\left(\frac{\theta-\Theta(\tau)}{\sigma(\tau)}\right).
\end{equation}
Here ${\cal C}$ is the ratio between the experimental count rate and the macroscopic scattering cross section of the sample, $2\theta$ is the scattering angle, $\tau$ indicates distinct reciprocal lattice vectors, $2\sigma(\tau)$ is the standard deviation for $2\theta$ near the Bragg peak at scattering angle $2\Theta(\tau)$, and $g(x)=\exp (-x^2/2)/\sqrt{2\pi}$ is a unity normalized gaussian distribution. $I(\tau)$ is the $2\theta-$integrated powder cross section for $|\vec{\tau}|=\tau$, which can be expressed as:
\begin{equation}
I(\tau)=\frac{4\pi^2\tan\Theta}{v_0\tau^3} N\sum_{|\vec{\tau}|=\tau}|F(\vec{\tau})|^2.
\end{equation}
Here $v_0$ is the unit cell volume, $N$ is the number of unit cells in the sample, and  $F(\vec{\tau})$ is the unit cell structure factor including Debye Waller factors.

Thermal expansion gives rise to a small shift in the scattering angle that can be written as follows:
\begin{equation}
\Delta\Theta=\frac{\Delta\tau}{\tau}{\tan \Theta},
\end{equation}
Neglecting changes in the intensity of nuclear Bragg peaks, the resulting difference intensity is given by
\begin{eqnarray}
\Delta {\cal I}(\theta) &=& \frac{\partial {\cal I}(\theta)}{\partial \Theta}\Delta\Theta\\
&=&{\cal C}\sum_{\tau}\frac{I(\tau)}{2\sigma(\tau)} g^{\prime}\left(\frac{\theta-\Theta(\tau)}{\sigma(\tau)}\right)\nonumber\\
&&\times\left(-\frac{1}{\sigma(\tau)}\right) \frac{\Delta \tau}{\tau} \tan \Theta (\tau).
\label{dI}
\end{eqnarray}
Here $g^{\prime}(x)=-xg(x)$. In general we have
\begin{equation}
\frac{\Delta\tau}{\tau}=\sum_i\frac{\partial \ln \tau}{\partial a_i}\Delta a_i,
\end{equation}
where the summation is over all temperature dependent lattice parameters including angular variables. In the present case of persistent tetragonal symmetry we have
\begin{equation}
\frac{\Delta\tau}{\tau}=-\left[ (1-(\hat{\tau}_{\parallel})^2)\frac{\Delta a}{a} +(\hat{\tau}_\parallel)^2\frac{\Delta c}{c}\right]
\label{dtot}
\end{equation}
where $\hat{\tau}_\parallel$ is the projection of the unit vector $\hat{\tau}$ on the tetragonal c-axis.
Having determined $\sigma(\tau)$ and $I(\tau)$ from Rietveld refinement in the paramagnetic phase, we then Rietveld refined the  difference between the low and high $T$ data with a functional consisting of the magnetic diffraction profile plus Eq.~\ref{dI} allowing only for adjustable thermal expansion coefficients $\Delta a/a$ and $\Delta c/c$ (Eq.~\ref{dtot}). The magnetic diffraction profile was developed using representation analysis as implemented in \textit{SARAh} \cite{sarah} and \textit{FULLPROF} \cite{fullprof}.

\subsection{Electronic Structure}

Band structure calculations were performed using the full-potential linearized augmented plane wave (FP-LAPW) method implemented in the WIEN2K package \cite{wien2k}. The Perdew, Burke, Ernzerhof version of the generalized gradient approximation \cite{pbe-gga} (PBE-GGA) was used for the exchange correlation potential and an onsite Coulomb repulsion of 5 eV was added to Ce sites to approximately account for electronic correlations in the narrow 4f bands\cite{alyahyaei,andersson}. Because the Ce states do not contribute to the electronic states at the Fermi level, spin-orbit coupling was not included in the calculation. The lattice parameters and atomic positions used for the calculation were taken from the neutron diffraction refinement at T = 300 K and are listed in Table \ref{ST300}.

\section{Experimental Results}

\label{results}

\subsection{Chemical Structure}

\label{chem_el_structure}

\begin{figure}[t!]
\includegraphics[width=1.0\columnwidth,clip]{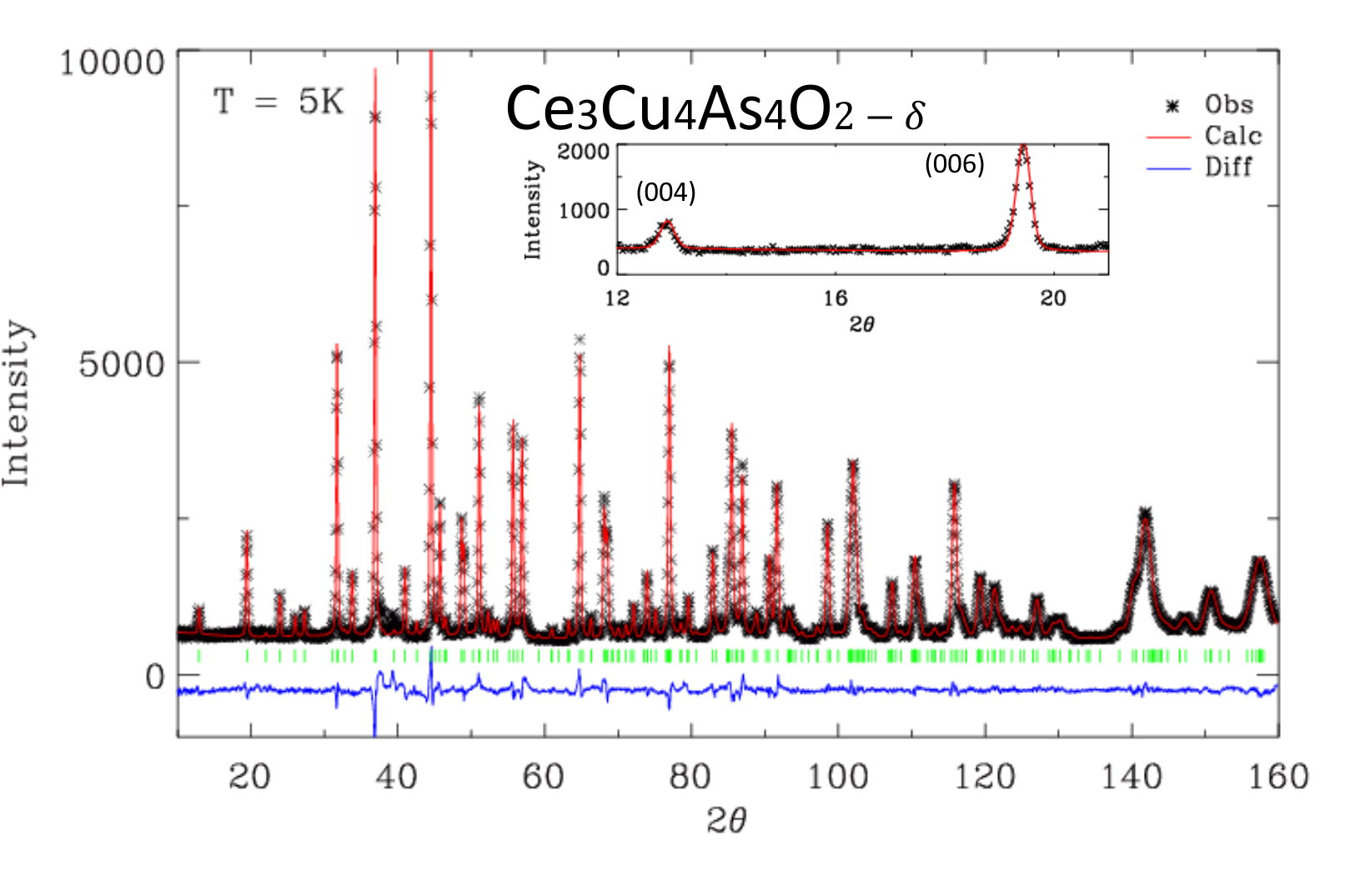}
\caption{\label{structurefit5K} Rietveld analysis of neutron diffraction from $\rm Ce_3Cu_4As_4O_2$ powder at T = 5 K. The data were acquired on BT1 at NIST with $\lambda = 1.5398 \AA$, with a Ge(311) monochromator  and 60$^\prime$ in-pile collimation. The red line shows the GSAS fit \cite{gsas}. The blue line shows the difference between model and diffraction data, while the green vertical lines show the calculated peak positions.The insert shows the detailed fit near the (004) and (006) nuclear Bragg peaks. All peaks are resolution limited and the Rietveld fit provides an acceptable account of the data. }
\end{figure}

\begin{figure}
\includegraphics[width=1.0\columnwidth,clip]{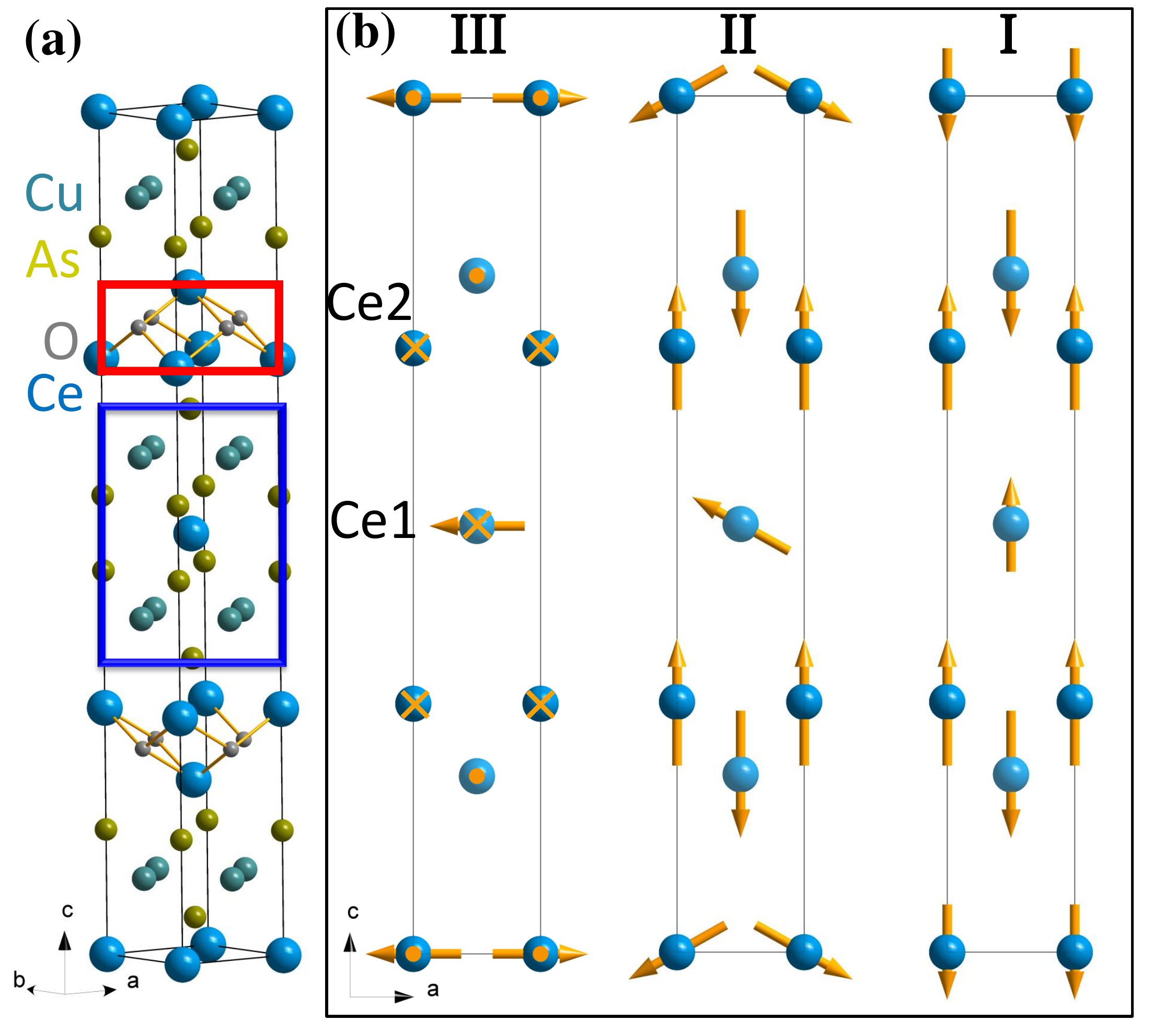}
\caption{\label{structure} (a) Ce$_3$Cu$_4$As$_4$O$_2$ crystal structure with space group \textit{I 4/mmm}. (b) Spin structures proposed for the magnetic phases \textit{I}, \textit{II} and \textit{III}. In frame (b), the arrows represent the spin direction and relative length. The spin structures for phase \textit{I} and \textit{III} can be determined directly from neutron diffraction Bragg intensities, while that for phase \textit{II} is proposed based on the T - and Q - dependence of inelastic magnetic scattering. More details are referred to the section IV (B).}
\end{figure}

The low temperature (T = 5 K) crystal structure was determined from the Rietveld refinement of the powder neutron diffraction data shown in Fig.~\ref{structurefit5K}. A single phase fit can account for the data with space group \textit{I4/mmm} and $\chi^2$ = 2.7. Powder X-ray diffraction data had indicated small oxygen defficiency in most reported $\rm R_3T_4As_4O_{2-\delta}$ compounds (T = Ni or Cu)\cite{jiakui}. However, the inset to Fig.~\ref{structurefit5K} shows the refinement of the oxygen stoichiometry in $\rm Ce_3Cu_4As_4O_{2-\delta}$ based on the present neutron data, which yields the best refinement (lowest $\chi^2$) for $\delta=0.00(5)$ and therefore indicates the compound is stoichiometric. No magnetic diffraction is visible in the BT1 data. This is however not inconsistent with the MACS data where the strongest  magnetic peak at T = 5 K has an integrated intensity of just 0.4\% of the strongest nuclear peak, which is too weak to be detected by BT1. A summary of the structural information is provided in table~\ref{ST300} and ~\ref{ST}.

\begin{table}[h!]

 \setlength{\tabcolsep}{8pt}
\caption{ T = 300 K atomic positions for Ce$_3$Cu$_4$As$_4$O$_2$ with space group \textit{I 4/mmm} determined by the Rietveld analysis\cite{gsas} of neutron powder diffraction data. The corresponding reduced chi-square  measure of goodness of fit is  $\chi^2$ = 2.7. The room temperature lattice constants  are \textit{a} = 4.07733(10) \AA ~ and \textit{c} = 27.4146(9) \AA. }   \label{ST300}
  \begin{tabular}{ c || c| c | c | c | c  }
  \hline
Atom & Site & x & y & z & $U_{iso}(\AA)$  \\ \hline
 &  &  &  &  & \\
  Ce1 & 2a & 0 & 0 & 0 & 0.0088(6)\\
 &  &  &  &  & \\
  Ce2 & 4e & 1/2 & 1/2 & 0.2069(1) & 0.0088(6)\\
 &  &  &  &  & \\
   Cu &  8g & 1/2 & 0 & 0.0975(5) & 0.0133(4)\\
 &  &  &  &  & \\
   As1 &  4e & 0 & 0 & 0.1459(1) & 0.0067(4)\\
 &  &  &  &  & \\
   As2 &  4e & 1/2 & 1/2 & 0.0454(2) & 0.0067(4)\\
 &  &  &  &  & \\
O &  4d & 1/2 & 0 & 1/4 & 0.0051(5)\\\hline
    \end{tabular}
\end{table}

\begin{table}[h!]
 \setlength{\tabcolsep}{8pt}
\caption{T = 4.5 K Atomic positions for Ce$_3$Cu$_4$As$_4$O$_2$ in space group \textit{I 4/mmm}  determined by the Rietveld analysis \cite{gsas} of powder diffraction data. The corresponding reduced chi-square  measure of goodness of fit is  $\chi^2$ = 3.4. The corresponding lattice constants  are \textit{a} = 4.06344(9) \AA ~ and \textit{c} = 27.3365(7) \AA. }   \label{ST}
  \begin{tabular}{ c || c| c | c | c | c  }
  \hline
Atom & Site & x & y & z & $U_{iso}(\AA)$  \\ \hline
 &  &  &  &  & \\
  Ce1 & 2a & 0 & 0 & 0 & 0.0029(5)\\
 &  &  &  &  & \\
  Ce2 & 4e & 1/2 & 1/2 & 0.2068(1) & 0.0029(5)\\
 &  &  &  &  & \\
   Cu &  8g & 1/2 & 0 & 0.0973(4) & 0.0030(5)\\
 &  &  &  &  & \\
   As1 &  4e & 0 & 0 & 0.1462(7) & 0.0002(3)\\
 &  &  &  &  & \\
   As2 &  4e & 1/2 & 1/2 & 0.04504(7) & 0.0002(3)\\
 &  &  &  &  & \\
O &  4d & 1/2 & 0 & 1/4 & 0.0007(4)\\\hline
    \end{tabular}
\end{table}

Illustrated in Fig.~\ref{structure}(a), the structure of $\rm Ce_3Cu_4As_4O_2$ consists of  alternating layers of $\rm CeCu_4As_4$ (blue box) and $\rm Ce_2O_2$ (red box). In $\rm CeCu_4As_4$, Ce atoms lie at the vertices of a simple square lattice sandwiched between $\rm Cu_2As_2$ layers. This structural unit is found in the $\rm ThCr_2Si_2$ structure, which is familiar from 122 iron superconductors such as $\rm BaFe_2As_2$ where Ba occupies the site equivalent to Ce in the $\rm CeCu_4As_4$ layer \cite{su}. There are also a number of heavy fermion systems with a magnetic ion on this site including $\rm CeT_2Si_2$ \cite{shuzo}$^,$\cite{stockert}$^,$\cite{regnault}$^,$\cite{dijk} (T=Pd, Ru, Rh, Cu) and $\rm URu_2Si_2$ \cite{collin}.

\begin{figure}[t!]
\includegraphics[width=\columnwidth,clip,angle=0]{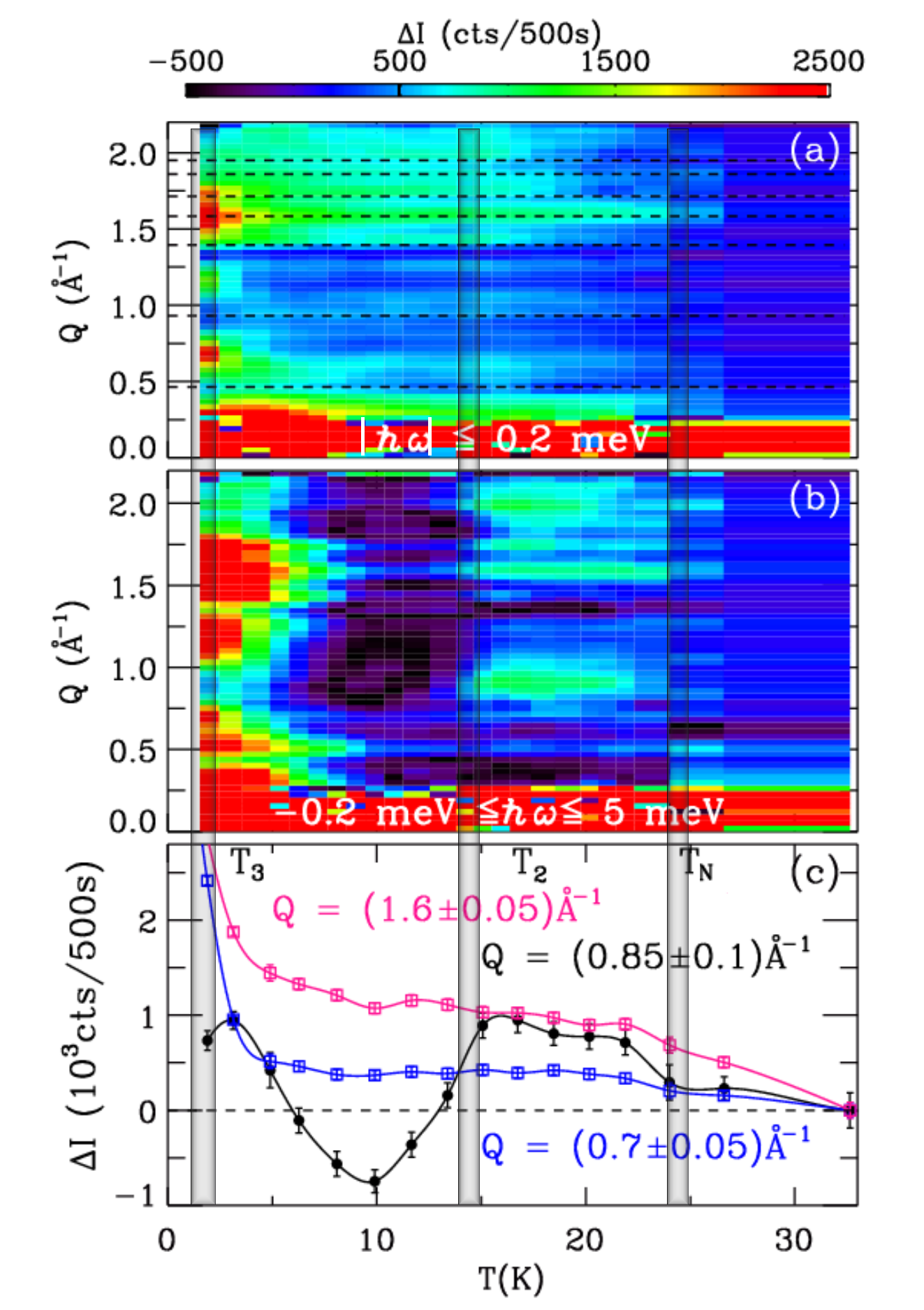}
\caption{\label{Neutron2} Color image of (a) elastic and (b) energy integrated neutron scattering versus temperature and wave vector. The data were acquired on the MACS instrument at NIST using the spectroscopic and energy integrating detectors respectively. Corresponding data sets acquired at $T=32$~K were subtracted from each to expose small temperature dependent effects. The incident neutron energy was 5 meV. Gray vertical regions indicate temperature intervals within 5\% of the peak maxima in the magnetic specific heat\cite{jiakui}. Horizontal dashes lines indicate the locations of allowed nuclear Bragg peaks. (c) Temperature dependence of the average scattering intensity in select ranges of wave vector transfer. Black symbols are from the data presented in frame (b). Red and blue symbols are from (a). In all the figures where data are from neutron scattering experiments, the error bars indicate one standard deviation.}
\end{figure}

In the $\rm Ce_2O_2$ layer, a 45$^0$ rotated square lattice of O with lattice parameter $a/\sqrt{2}$ is sandwiched by square lattices of Ce with lattice parameter $a$, such that half  of the squares of the O lattice have Ce above and the other half have Ce below. This structural element is also found in the 1111 type Fe superconductors and specifically in CeFeAsO\cite{zhao} and CeNiAsO\cite{luo}.

\subsection{Magnetic Neutron Scattering}
\label{mag_neutron_results}

\subsubsection{Temperature dependence}

Three peaks in the specific heat of polycrystalline $\rm Ce_3Cu_4As_4O_2$ were previously reported and associated with magnetic transitions\cite{jiakui}.  In order of decreasing temperature, we shall denote the corresponding phases by I ($T_{2}<T<T_{N}$), II ($T_{3}<T<T_{2}$), and III ($T<T_{3}$) respectively. In search for the associated atomic scale magnetic correlations, we acquired neutron diffraction data for temperatures between 1.5 K ($<$ T$_3$) and 40 K ($>$ T$_N$) using a high intensity configuration of the MACS spectrometer. For sensitivity to weak temperature dependent scattering we subtract a high statistics data set acquired at $T=32$~K and plot the difference data as a color image in Fig.~\ref{Neutron2}. The MACS instrument offers simultaneous energy resolved data and final energy integrated data shown respectively in frames Fig.~\ref{Neutron2}(a) and Fig.~\ref{Neutron2}(b). Coincident with each of the transition regimes inferred from specific heat data (vertical grey regions) are anomalies in the temperature dependent neutron scattering data. Fig.~\ref{Neutron2}(c) provides a detailed view of the temperature dependent intensity averaged over relevant ranges of wave vector transfer.

Below $T_{N}$ we observe the development of Bragg diffraction for 1.5~$\AA^{-1}<Q<1.9~\AA^{-1}$ in Fig.~\ref{Neutron2}(a). In being associated with a net increase in intensity, these features in the ``thermo-diffractogram" are distinguished from the effects of thermal expansion, which produce matched minima and maxima surrounding each nuclear Bragg peak (thin dashed lines).

Fig.~\ref{Neutron2}(b) and Fig.~\ref{Neutron2}(c) (black symbols and line) show a distinct reduction in inelastic scattering for temperatures below the characteristic temperature $T_{2}$ extracted from bulk properties. This observation was reproduced in experiments using both a flow cryostat and a dilution fridge, and the temperature and momentum regimes are inconsistent with contributions from cryogenic fluids such as nitrogen or helium. We therefore associate this feature with a reduction in inelastic magnetic scattering for $T<T_{2}$. The total scattering sum-rule requires that reduced inelastic scattering appears elsewhere in ${\cal S}(Q\omega)$. One option is a shift of all the lost spectral weight beyond the 5 meV cut-off of the experiment (see Eq.~\ref{energyintegrated}).Alternatively a reduction in inelastic scattering can be associated with the development of elastic scattering and static correlations.  In that case a possible explanation for the  absence of elastic magnetic scattering associated with $T_{2}$ (see Fig.~\ref{Neutron2}(a)) is that the associated component in ${\cal S}(Q\omega)$ is extinguished or severely weakened by the so-called polarization factor, which removes scattering associated with spin components parallel to wave vector transfer. This would be the case for a modulated in-plane spin structure in phase II where the staggered moment is oriented along the characteristic wave vector of the magnetic structure.

\begin{figure}[t!]
\includegraphics[width=\columnwidth,clip,angle=0]{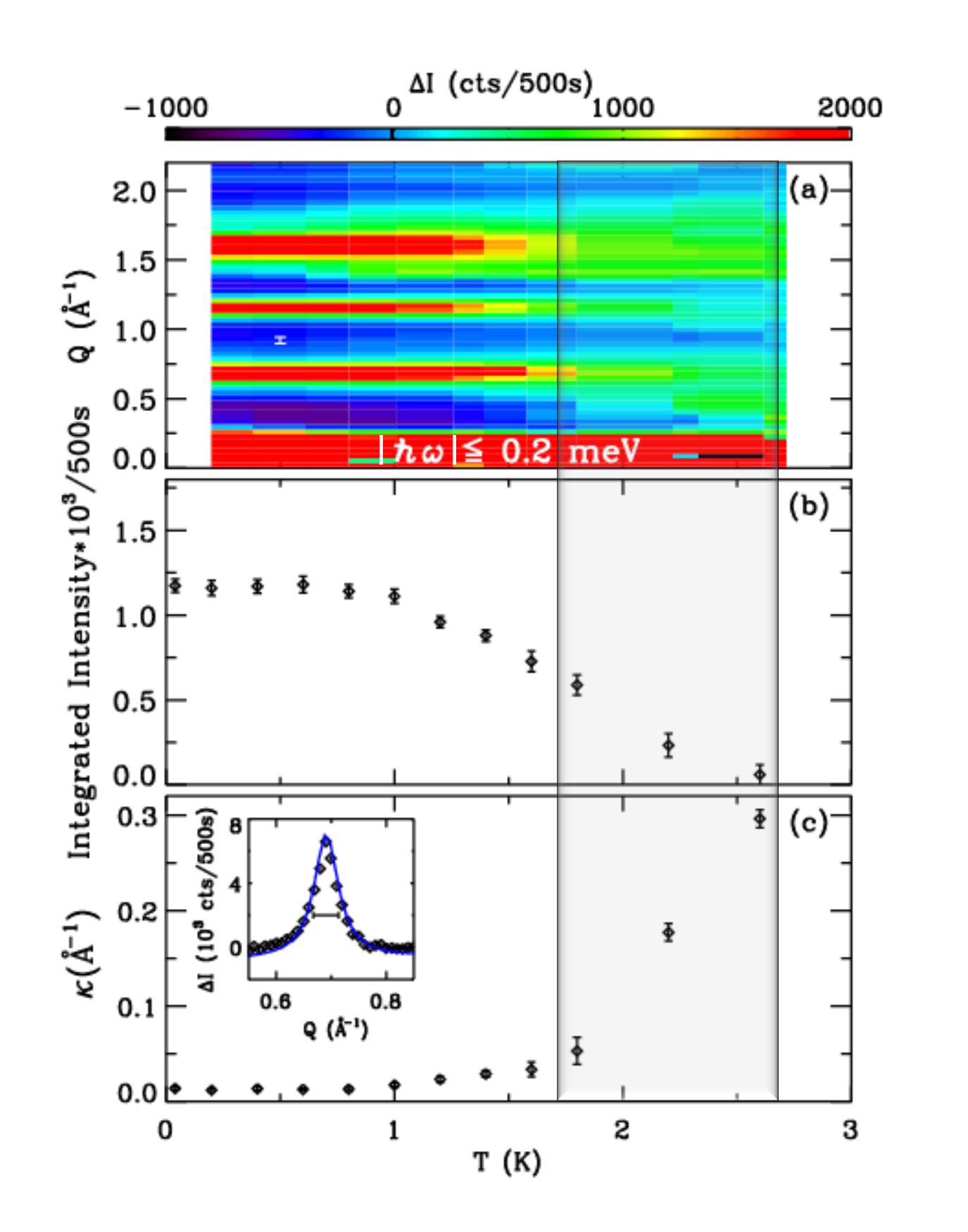}
\caption{\label{Neutron4} (a) Image of elastic neutron scattering versus temperature and wave vector transfer in the low temperature regime. Data collected at T = 15 K has been subtracted to remove scattering intensities of nuclear Bragg peaks. (b) Intensity and (c) inverse correlation length versus temperature extracted from fits of a Lorentzian convoluted with a Gaussian of the resolution width (horizontal bar within inset) to the most intense magnetic peak. An example of such a fit is shown in the inset to (c).}
\end{figure}

The clearest anomaly in scattering is associated with the lowest temperature transition. The temperature regime for $T<T_{3}$ (phase III) is detailed in Fig.~\ref{Neutron4}. Phase III is associated with the development of strong Bragg peaks and we shall argue that only in this phase is the spin direction perpendicular to the direction of modulation so the polarization factor does not extinguish the low $Q$ peaks. Fig.~\ref{Neutron4}(b) shows the integrated intensity, that develops in an order parameter like fashion for $T<T_{3}$ and Fig.~\ref{Neutron4}(c) shows the half width at half maximum of the Lorentzian fit (inset to Fig. \ref{Neutron4}(c)). After correcting for resolution effects the low T limit corresponds to a correlation length above 75 $\AA$.

While we have identified distinct anomalies in the $T$ dependence of neutron scattering at each transition, all except perhaps those associated with $T_{3}$ are unusually broad in temperature compared to a standard second order phase transition observed through diffraction with a cold neutron beam. This mirrors the specific heat anomalies\cite{jiakui} and suggests we are dealing with cross-over phenomena rather than critical phase transitions, which could result from disorder.

\begin{figure}[t!]
\includegraphics[width=1.0\columnwidth,clip]{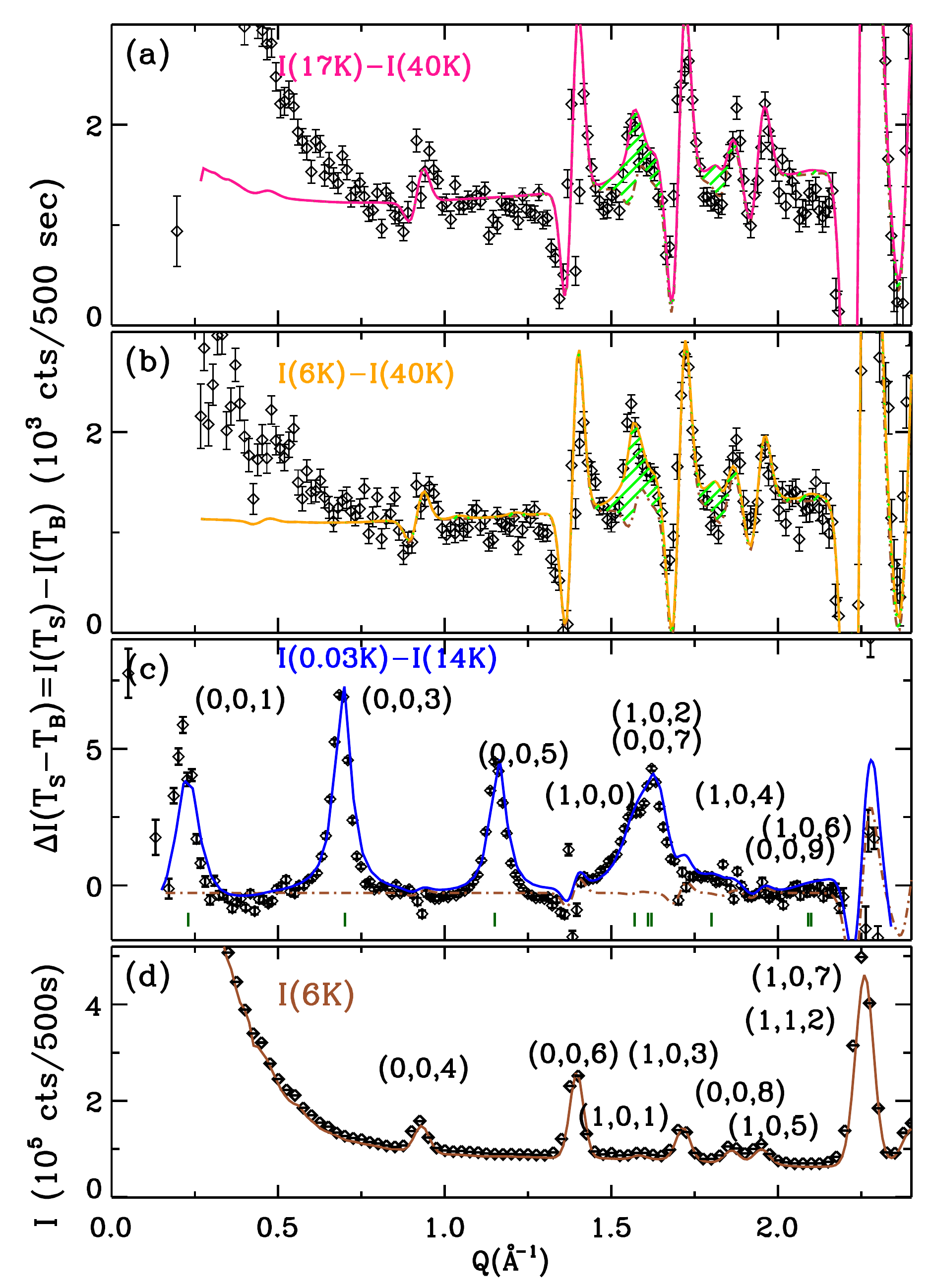}
\caption{\label{diffraction_pattern}  Neutron diffraction patterns for Ce$_3$Cu$_4$As$_4$O$_2$ obtained at MACS with T = (a) 17 K, (b) 6 K and  (c) 30 mK. The fitted lines in (a)-(c) include a conventional Rietveld model for the magnetic scattering plus the peak derivative profile (dashed line in (c)) described in Section II(B). The pattern in panel (b) includes the calculated diffraction pattern corresponding to the proposed spin structure in Phase \textit{II} (Fig. \ref{structure}) with moment size listed in table \ref{summary}. Such contribution would go undetected experimentally due to the polarization factor when spins are almost parallel to the wave vector. The line in (c) is a Rietveld fit to the structural peaks.}
\end{figure}

\subsubsection{Diffraction in each phase}

To elucidate the nature of each magnetic phase, we  isolate the $Q$ dependence of the associated scattering by subtracting data collected for $T>T_{N}$ from high statistics data acquired in each phase.

\begin{figure}[t!]
\includegraphics[width=0.95\columnwidth,clip,angle=0]{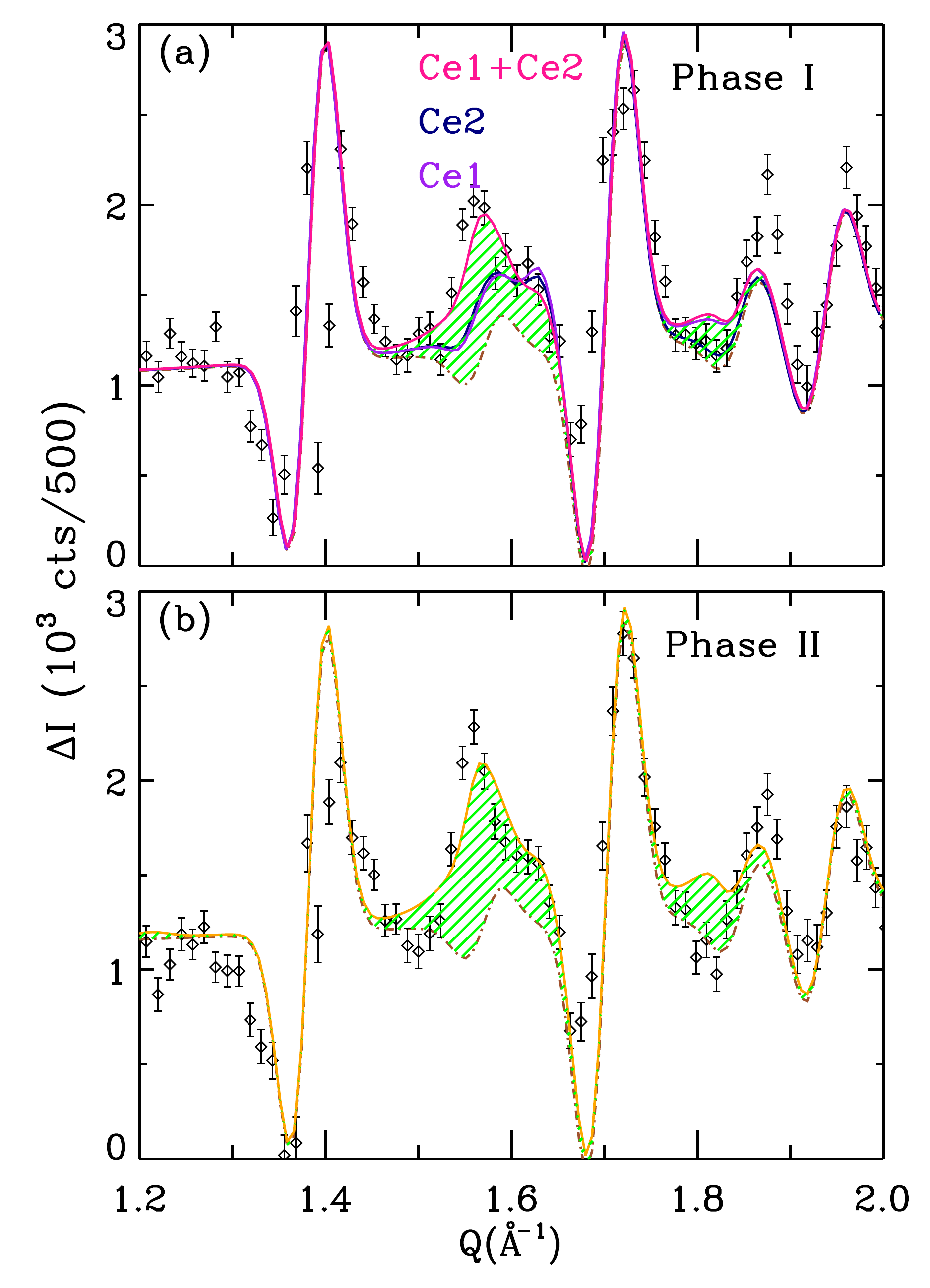}  
\caption{\label{detailph} Detailed Q-dependence of neutron scattering data characterizing the magnetic order in (a) phase I ($16K < T < 24K$) and (b) phase II ($2K < T < 16K$). A background data set acquired at $T=40$~K was subtracted to expose temperature dependent features associated with the phase transitions. The lines through the data include the peak derivative features associated with thermal expansion (Eq.~\ref{dI} and the calculated magnetic diffraction associated with the structures proposed in Fig.~\ref{structure}).
}
\end{figure}

\begin{figure}[t!]
\includegraphics[width=1.0\columnwidth,clip]{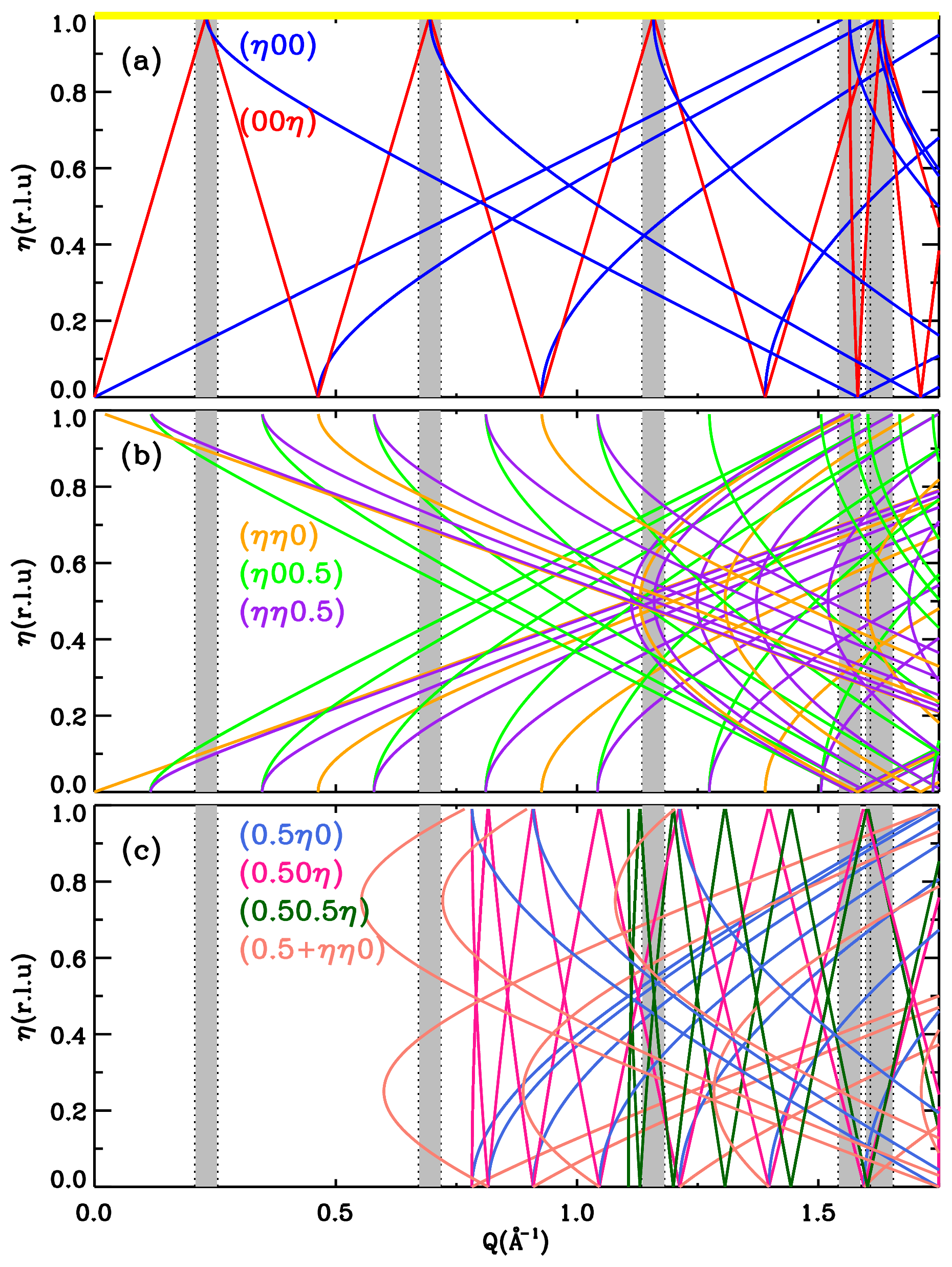}
\caption{\label{kvectors} Relationship between high symmetry magnetic modulation wave vectors distinguished by colors and the location of magnetic satellite peaks in the powder diffraction pattern for $\rm Ce_3Cu_4As_4O_2$. The vertical grey regions show the locations of the experimentally observed magnetic Bragg peaks. The low $T$ phase III has all indicated Bragg peaks whereas only the peaks near 1.5 $\AA^{-1}$ are visible in Fig.~\ref{detailph} for phases I-II.  The calculated wave vector transfer associated with the horizontal yellow line in (a) ${\bf k}_1={\bf c}^*$ or equivalently ${\bf k}_1={\bf a}^*$ are consistent with the observed  magnetic Bragg peaks. Frames (b) and (c) show the remaining singly indexed high symmetry magnetic wave vectors that fail to account for the magnetic Bragg peaks observed in phase III.
}
\end{figure}

For phase I the difference pattern is shown in Fig \ref{diffraction_pattern} (a). Apart from the characteristic zig-zag anomalies associated with thermal expansion, two distinct Bragg peaks are associated with Phase I. A detailed view of these peaks is provided in Fig.~\ref{detailph}. Discounting peak derivative line shapes indicated with the dashed lines described by Eq.~\ref{dI}, we identify two  peaks (shadowed in bright green lines)  whose intensities are less than $1\%$ of nuclear ones at wave vector transfer $Q\sim 1.57 ~ \AA^{-1}$ and $Q\sim 1.63 ~ \AA^{-1}$. Nearby there are no strong nuclear peaks and is therefore possible to obtain reliable magnetic diffraction.The additional intensity at $Q\sim 1.8 \AA^{-1}$ could be misleading due to nuclear subtraction. The figure shows  the peaks are resolution limited and this sets a low T limit of  75 \AA\ on the correlation length associated with translational symmetry breaking. The position of the peaks is indicated by the two vertical hatched regions in Fig.~\ref{kvectors} for $Q>1.5 \AA^{-1}$. Assuming that a single propagation vector is associated with Phase I, Fig. \ref{kvectors} shows that this must be $\textbf{k}_1\equiv (0, 0, 1.0(2))$. The absence of magnetic Bragg peaks of the form $(0, 0,2n+1)$, where $n$ is an integer, indicates these peaks are extinguished by the polarization factor and so calls for spins polarization along $\bf c$.

\begin{figure}[t!]
\includegraphics[width=1.0\columnwidth,clip]{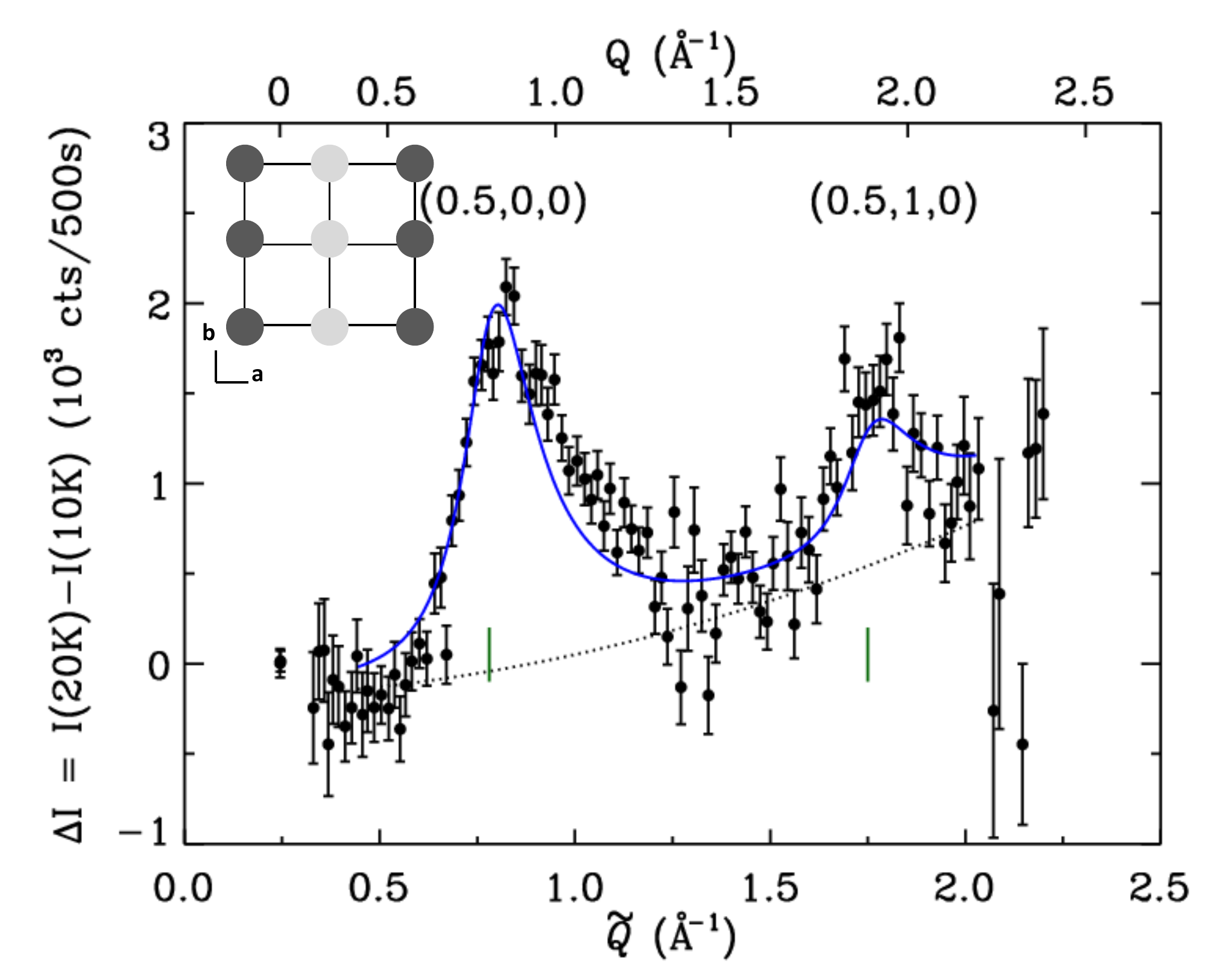}
\caption{\label{Neutron3} $\tilde{Q}$-dependence difference intensity 20~K and 10~K from the energy integrating detectors on the MACS spectrometer. The data is plotted as a function of wave vector transfer $\tilde{Q}(\theta,\omega)$ for energy transfer $\langle\hbar\omega\rangle=1.2(3)$~meV inferred from the fitting analysis (see section~\ref{analysis}). The upper horizontal axis shows wave vector transfer for elastic scattering. The solid blue line is a Warren-like line shape for quasi-two-dimensional AFM correlations with characteristic wave vector ${\bf k}_2=(1/2,0,0)$ and in-plane dynamic correlation length $\xi=8.2(6)$~\AA. The dashed line is a $Q^2$ background ascribed to the difference in the Debye Waller factor for incoherent scattering at the two temperatures involved. The inset shows the pattern of quasi-two-dimensional fluctuation associated with ${\bf k}_2$, with black and grey spots representing alternative spin . The proposed ordered state for phase II has spins parallel to ${\bf k}_2$ (Fig.~\ref{structure}(b)).}
\end{figure}

The diffraction pattern for Phase II is shown in Fig \ref{diffraction_pattern} (b) with the detail in Fig.~\ref{detailph} (b). Surprisingly there is virtually no change in the position or widths of the Bragg peaks observed for Phase I. The black symbols in Fig. 3(c) however, show that Phase II is associated with the loss of inelastic scattering. To better understand the associated dynamic correlations, Fig.~\ref{Neutron3} shows the $Q-$dependence of the energy integrated scattering that is lost upon cooling into phase II. It takes the form of asymmetric peaks, which, in the direction of increasing $Q$, rise more abruptly than they decay. This Warren-like line shape\cite{warren} is a clear indication of low dimensional correlations. We shall later show the correlations to be quasi-two-dimensional. In that case rods of scattering extend perpendicular to the plane of long range correlations. The sharp leading edge is associated with these rods becoming tangents to the Ewald sphere and the long trailing tail results from the fraction of the Ewald sphere pierced by that rod decreasing in inverse proportion to the area of the sphere.

From Fig.~\ref{Neutron3} we see that the characteristic wave vector of the rod is 0.8 $\AA^{-1}$. The proximity of this number to $a^*/2=0.77 ~ \AA^{-1}$ indicates that the spin fluctuations of phase II are associated with doubling the unit cell in the basal plane. Possible explanations for the absence of a corresponding magnetic Bragg peak in Fig.~\ref{detailph} (c) are (1) that a competing instability prevails so that static long range AFM correlations of the ${\bf k}_2={\bf a}^*/2$ variety never materialize and (2) that a polarization factor extinguishes the magnetic Bragg peaks associated with this order in the low $Q$ regime where the experiment has sufficient sensitivity to detect them.

Shown in Fig.~\ref{diffraction_pattern} (c), the strongest magnetic diffraction peaks are associated with Phase III. As indicated in Fig.~\ref{kvectors} all observed peaks associated with phase III can be accounted for by a wave vector ${\bf k}_1={\bf c}^*$ or equivalently ${\bf k}_1={\bf a}^*$. The appearance of Bragg peaks of the form $(0 0,2n+1)$ in Phase III indicate spin components within the basal plane and a spin-flop type transition relative to phases I and II. To be quantified in section \ref{analysis}, the magnetic peaks are slightly broader than the nuclear peaks even at 30 mK indicating an element of disorder in the magnetic structure or strong thermal diffuse magnetic scattering with a length scale shorter than the  coherence length of the chemical structure. Upon heating, the intensity of the magnetic peaks decreases precipitously above 2 K, though a broad feature remains at all magnetic Bragg positions up to T = 2.6 K. This indicates dynamic diffuse scattering near the critical temperature of the spin-flop transition.

\subsection{Physical properties}

\subsubsection{Magnetization}

\begin{figure}[t!]
\includegraphics[width=1.0\columnwidth,clip]{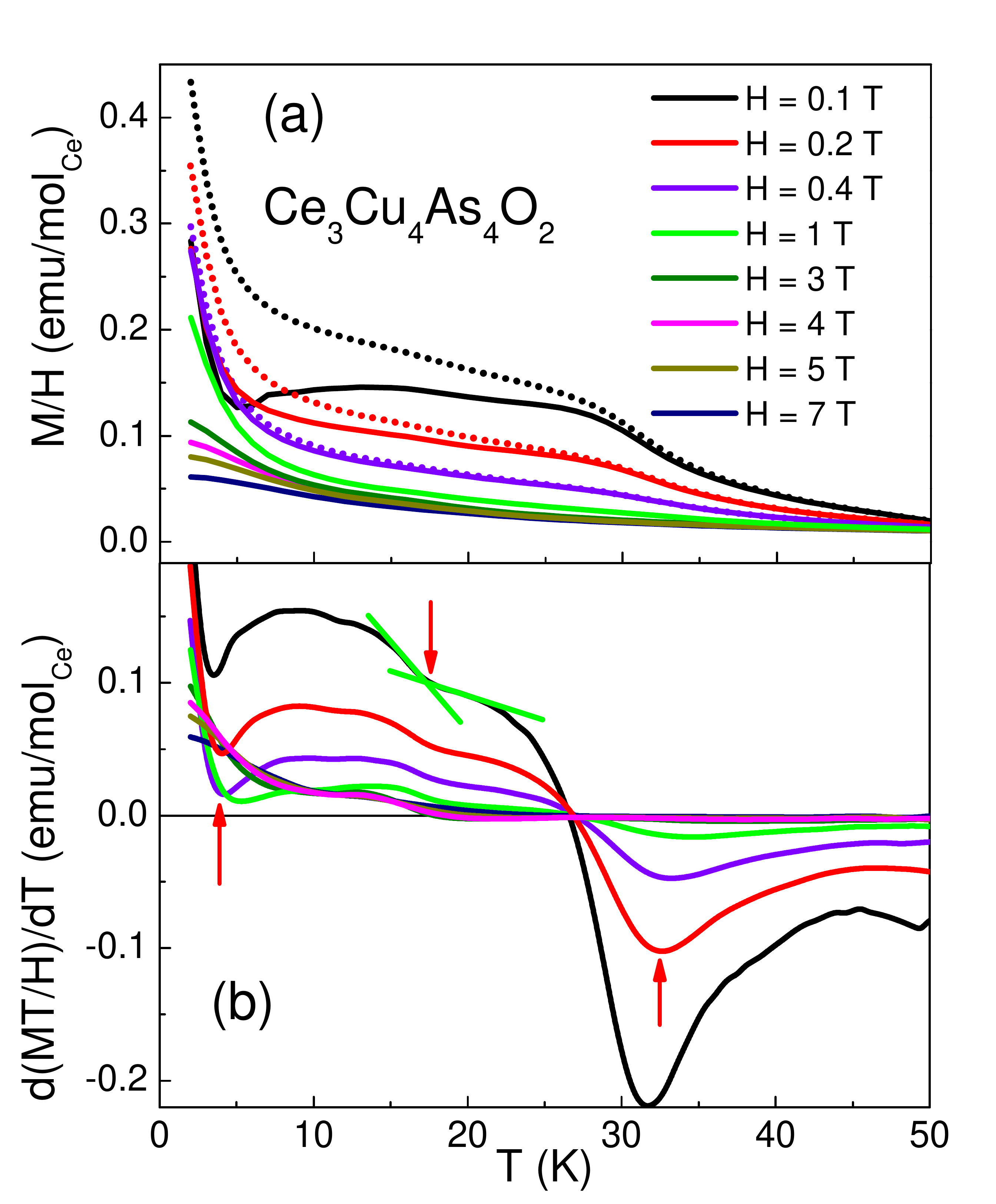}
\caption{\label{CeCu-MT} (a) ZFC (solid lines) and FC (dotted lines) temperature-dependent M/H measured in applied magnetic fields up to H = 7 T. (b) The temperature derivatives of M/H$*$T for different magnetic fields. The three transition temperatures are marked by vertical arrows for H = 0.}
\end{figure}

The magnetization versus temperature was measured in different fields and the corresponding data are shown in Fig. \ref{CeCu-MT}(a). All magnetization measurements were conducted on polycrystalline samples, so the data represent the spherical average of the longitudinal response to the magnetic field $H$.

For temperatures below $\sim$ 30 K, the bifurcation between zero field cooled (ZFC) (solid lines) and field cooled (FC) (dashed lines) magnetization data occurs for fields up to H = 0.4 T. To identify potential magnetic  transitions, Fig. \ref{CeCu-MT}(b) shows the derivatives $H^{-1}d(MT)/dT$ \cite{fisher}, for which peaks are expected at magnetic phase transitions. Indeed, anomalies in $H^{-1}d(MT)/dT$, marked by vertical arrows, occur near each of the transitions previously identified in specific heat data \cite{jiakui}.  As the field increases, $H^{-1}d(MT)/dT$ data (Fig. \ref{CeCu-MT}(b)) indicate that the lowest transition temperature T$_{3}$ increases and reaches $\approx$ 7 K for H = 7 T, while the upper two transitions at $T_{N}$ and $T_{2}$ are largely unaffected by $H~<~7$~T.

\begin{figure}[t!]
\includegraphics[width=1.0\columnwidth,clip]{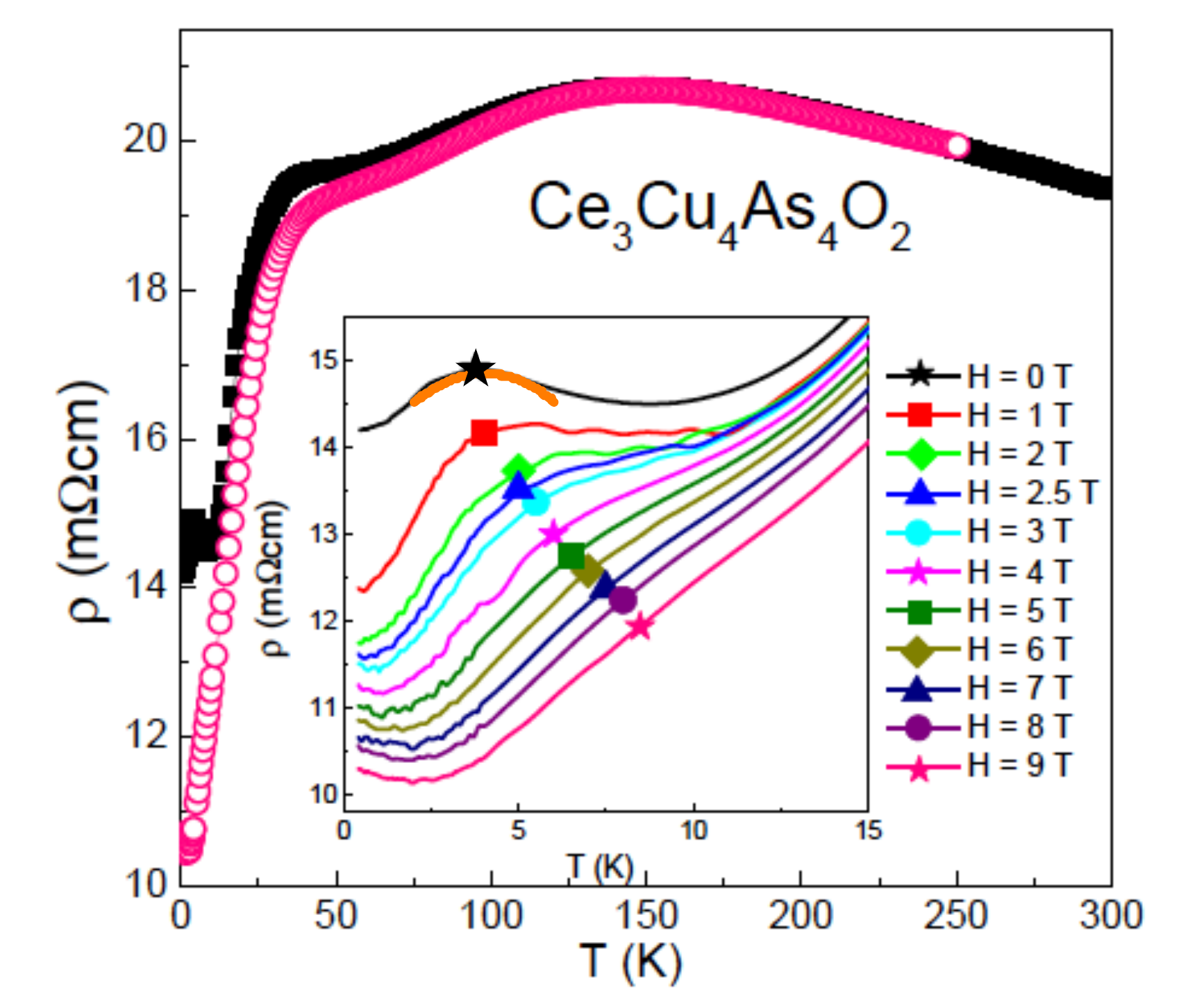}
\caption{\label{CeCu-R} The resistivity of Ce$_3$Cu$_4$As$_4$O$_2$ for H = 0 (squares) and H = 9 T (circles). Inset: enlarged view of the low temperature resistivity in this field range.  Solid orange line is an example of polynomial fits for H = 0 T to determine the transition temperature. }
\end{figure}

\subsubsection{Resistivity}
\begin{figure}[t!]
\includegraphics[width=1.0\columnwidth,clip]{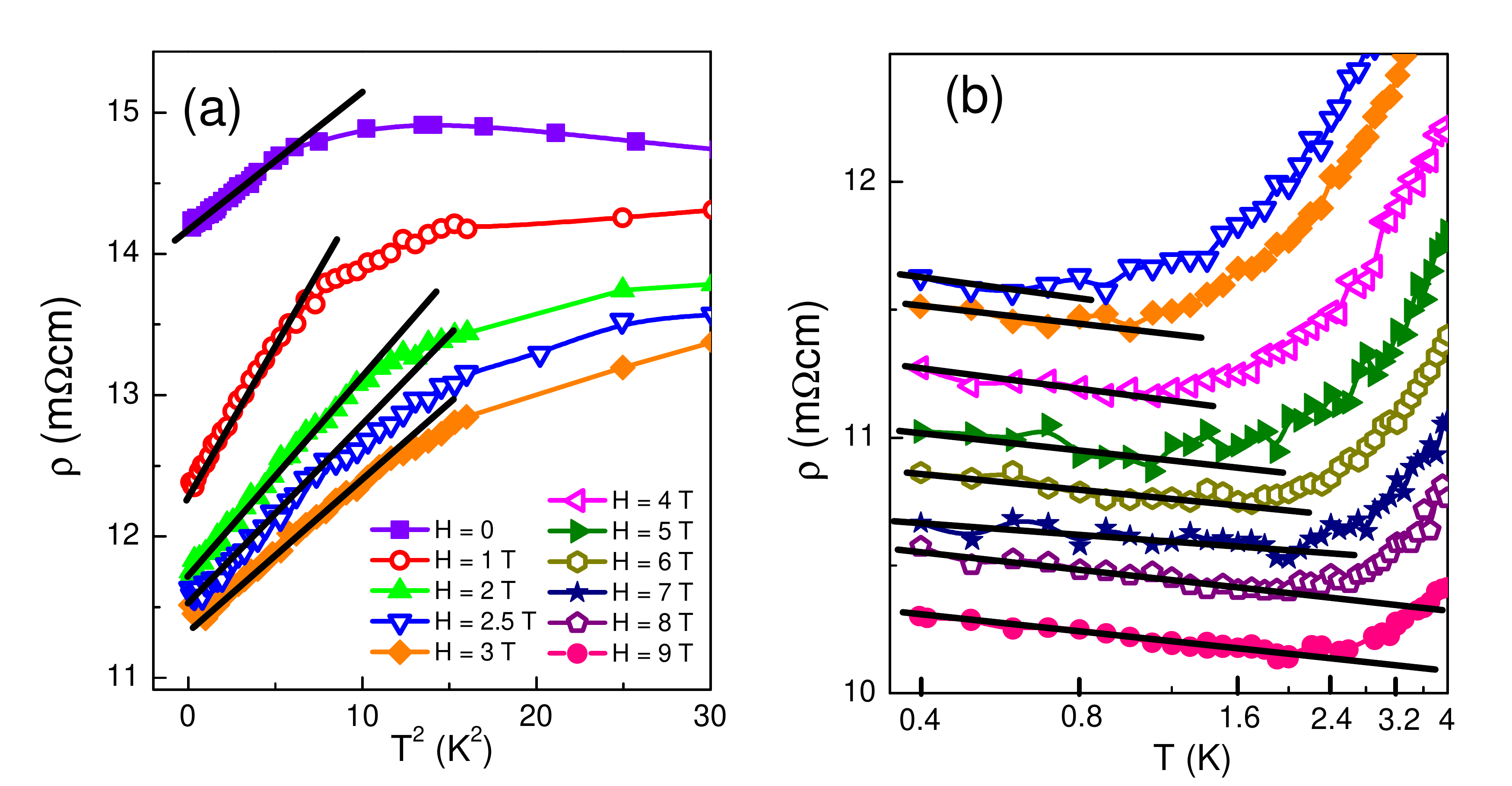}
\caption{\label{CeCu-NFL} (a): The quadratic dependence of resistivity for small magnetic fields between H = 0 T and H = 3 T. (b): The logarithmic dependence of resistivity for large fields between H = 2.5 T and H = 9 T.  }
\end{figure}

The field dependence of the resistivity measured on a dense polycrystalline sample is shown in Fig. \ref{CeCu-R}. The resistivity increases upon cooling below room temperature, reaching a maximum at T $\approx$ 150 K. The increase is typically associated with the single ion Kondo effect, with the maximum signaling the onset of inter-site coherence.  Below 150 K, poor metallic behavior is observed, as the resistivity decreases with temperature, albeit with large resistivity values $\rho > 10~{\rm m}\Omega$cm. As the temperature decreases below 50 K, a difference between  H = 0 (full symbols) and  H = 9 ~T (open symbols) data develops, well above the upper transition temperature $T_{N}$ inferred from specific heat data\cite{jiakui}. Notably however, the magnetoresistance reaches a maximum near $T_{3}$, as will be shown explicitly below. The inset of Fig. \ref{CeCu-R} shows the low temperature resistivity measured in different magnetic fields up to  H = 9 ~T. The resistivity decreases monotonically with increasing magnetic field. For  H = 0, a broad peak around 4 K is observed, which shifts to higher temperature with increasing field. In phase III, Fermi liquid (FL) behavior is evidenced by the quadratic temperature dependence of the resistivity (Fig. \ref{CeCu-NFL}a) up to H $\approx$ 3 T. This behavior is reminiscent of a Kondo lattice system, in which the competition between magnetic coupling and Kondo screening gives rise to a peak around the coherence temperature T$_{coh}$ and FL behavior below T$_N$\cite{schweitzer}. For fields beyond  H $\approx 2.5 - 3$ T however, a finite-T minimum in the resistivity develops, followed by a logarithmic increase of the resistivity upon cooling below $\approx$ 1 K as apparent in the semi-log plot in Fig. \ref{CeCu-NFL} (b). The $\rho \propto \log T$ behavior persists up to a temperature $T^*$ that increases with magnetic field.

The origin of the logarithmic temperature dependence of resistivity in Ce$_3$Cu$_4$As$_4$O$_2$ is not clear: either the formation of Kondo singlet \cite{hewson} or weak Anderson localization in a 2D systems \cite{anderson} could lead to $\rho(T) \propto \log T$ behavior, however neither phenomenon is relevant for Ce$_3$Cu$_4$As$_4$O$_2$. For the former scenario, a magnetic field breaks up the Kondo singlet, which is inconsistent with the observation that the logarithmic temperature dependence extends over a larger T-regime for higher magnetic fields. In the latter scenario, the magnetic field may tune the localization-delocalization transition. The upturn in resistivity could be observed in a system with weak localization\cite{giordano}. However, this localization usually increases the resistivity compared to the delocalized state. The magnetic field induced localization is not consistent with the observation of negative magnetoresistance in Ce$_3$Cu$_4$As$_4$O$_2$.

A complication in the interpretation of these data is that the polycrystalline nature of the sample implies the transport measurement is an average over all possible field and current directions with $ \bf H \parallel J$ in the tetragonal structure. For an anisotropic magnetic material such as Ce$_3$Cu$_4$As$_4$O$_2$ the high field low $T$ behavior could therefore be uncharacteristic and dominated by specific field directions.

\begin{figure}[t!]
\includegraphics[width=1.0\columnwidth,clip]{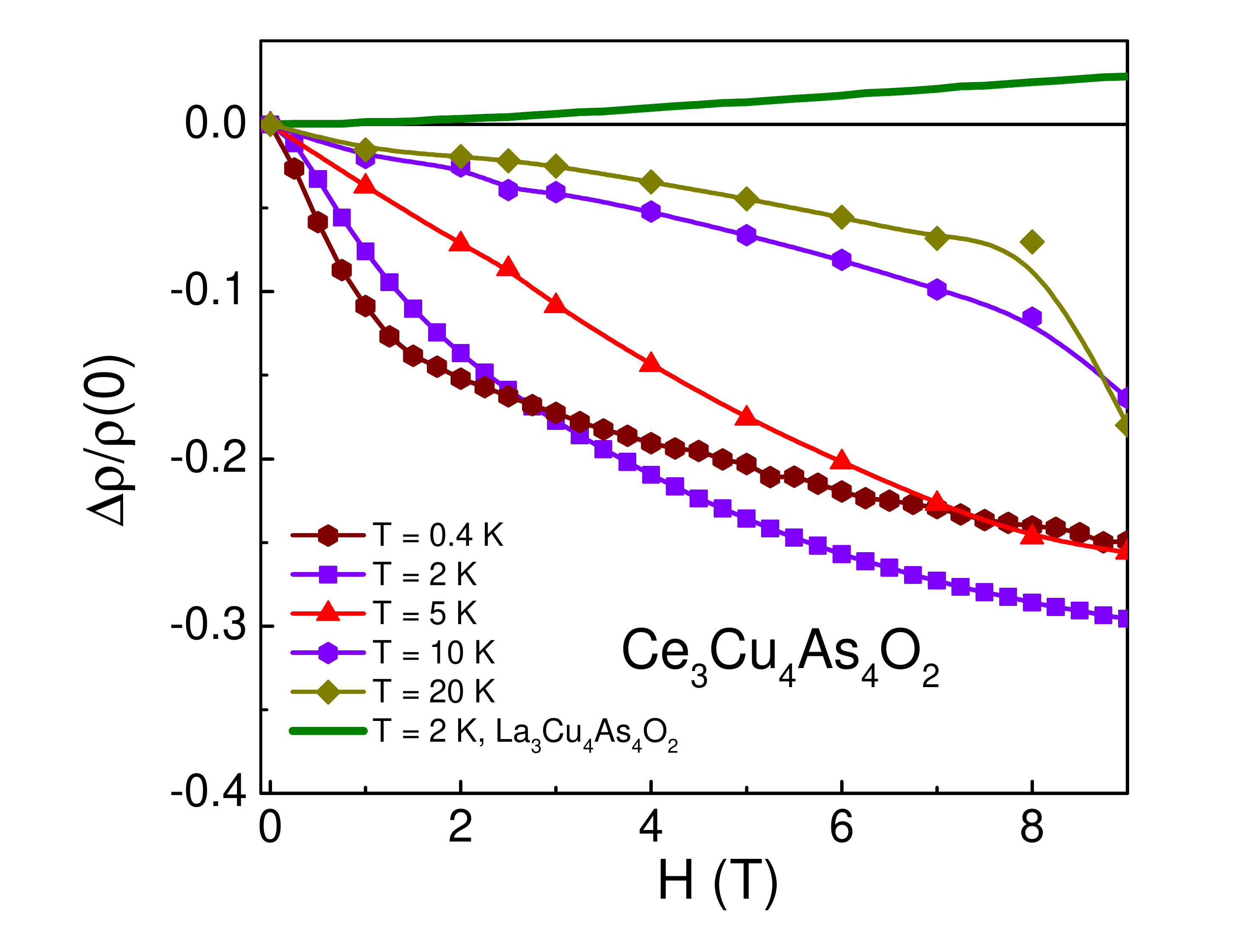}
\caption{\label{CeCu-MR} Magnetoresistance data for Ce$_3$Cu$_4$As$_4$O$_2$ (symbols) for various temperatures, with the magnetoresistance of non-magnetic La$_3$Cu$_4$As$_4$O$_2$ (line) at  T = 2 K shown for comparison.}
\end{figure}

The magnetoresistance is shown in Fig. \ref{CeCu-MR} for Ce$_3$Cu$_4$As$_4$O$_2$ (symbols) and the non-magnetic analogue La$_3$Cu$_4$As$_4$O$_2$ (line). For the La compound, the magnetoresistance is positive and small (less than 5\%), as expected for a normal metal in which the Lorentz force introduces electron scattering. For the Ce compound, the T = 2 K magnetoresistance is negative and reaches $\approx$ --30\% at H = 9 T. This suggests that the magnetism associated with the Ce$^{3+}$ ions is responsible for the large magnetoresistance in Ce$_3$Cu$_4$As$_4$O$_2$.

The magnetoresistance for Ce$_3$Cu$_4$As$_4$O$_2$ is largest at T = 2 K, close to T$_3$. Similar negative large magnetoresistance was also observed in the Kondo lattice systems Ce$_2$Ni$_3$Ge$_5$, CeB$_6$ and CeRhSn$_2$.\cite{steglich, woods, geibel} Above the magnetic ordering temperature, an important contribution to the resistivity is the Kondo scattering of conduction electrons off local moments. The magnetic field could effectively suppress the Kondo scattering and induce a negative magnetoresistance. This negative magnetoresistance will reach its maximum at the magnetic ordering temperature where the magnetic fluctuations dominate all other contributions to the resistivity.

\begin{figure}[t!]
\includegraphics[width=0.8\columnwidth,clip]{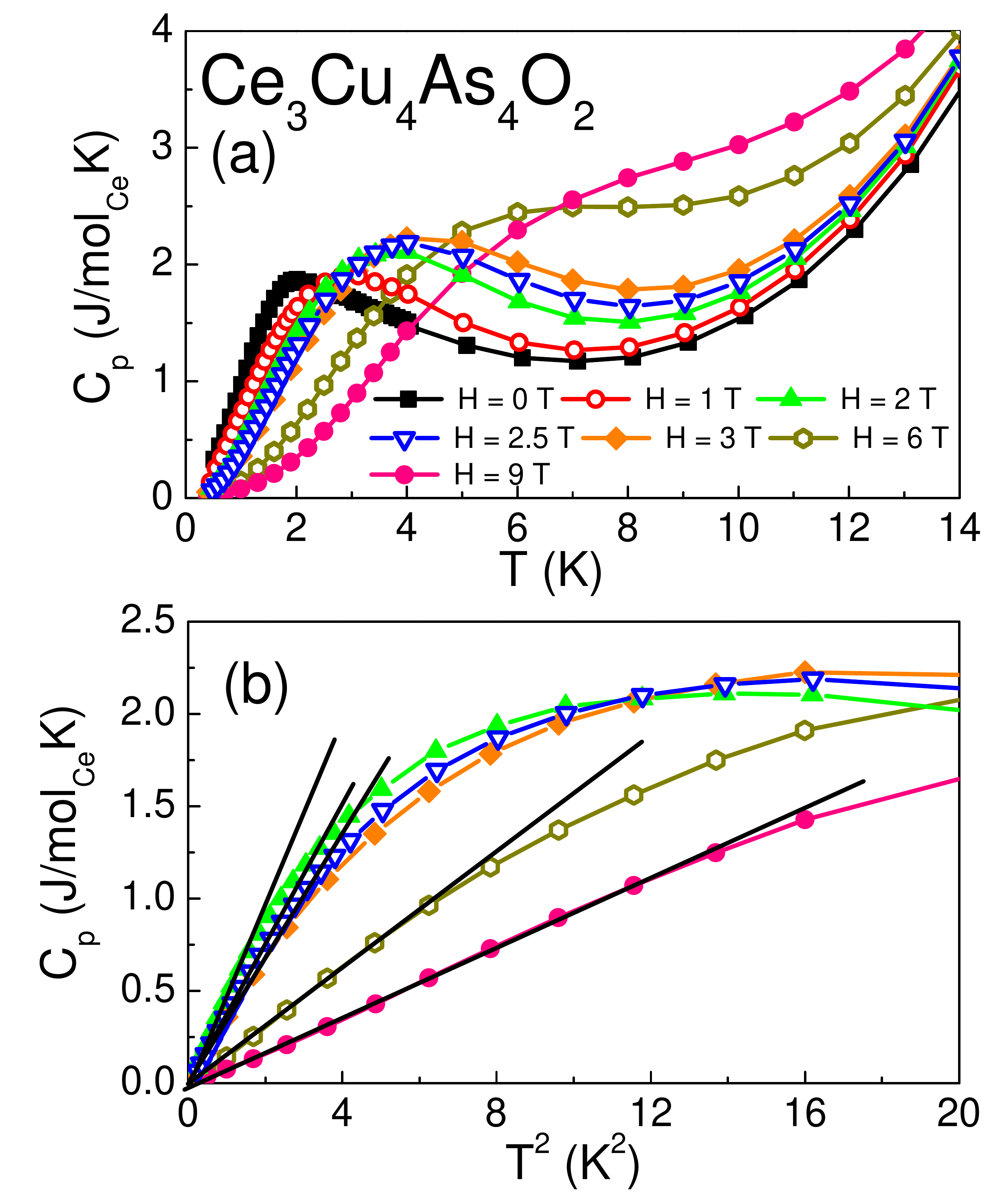}
\caption{\label{CeCu-Cp} (a) The specific heat of Ce$_3$Cu$_4$As$_4$O$_2$ measured from T = 0.4 K to T = 14 K, in applied magnetic fields between H = 0 and H = 9 T. (b) Quadratic temperature dependence of $C_p$ in magnetic fields between H = 2 T and H = 9 T. }
\end{figure}

\subsubsection{Specific Heat}

Additional information relevant to understanding the unusual low $T$ behaviors indicated by magnetization and transport measurements is provided by field-dependent specific heat data in Fig. \ref{CeCu-Cp}. From Fig. \ref{CeCu-Cp}(a), the broad peak associated with low $T$ magnetic ordering at $T_{3}~=~2.0$ K shifts to higher temperatures and broadens for lager fields applied to the polycrystalline sample. $C_p$ is quadratic in temperature for   H $~\geq~2.5$ T (Fig. \ref{CeCu-Cp}(b)). At H = 9 T, the quadratic dependence extends up to T = 4 K. This behavior can be associated with gapless excitations with a quasi-two-dimensional linear dispersion relation as in a two dimensional antiferromagnet. This indication of reduced dimensionality is consistent with the tube-like Fermi surface inferred from DFT (see section IV (C)). In zero field the low $T$ state indicated in Fig.~\ref{structure} would be expected to have an excitation gap similar in magnitude to $T_{3}$. The high field $C_p(T)~\propto~T^2$ regime might be associated with specific field directions in the polycrystalline experiment that are transverse to the easy axes and induce gapless behavior.

\section{Analysis}
\label{analysis}

\subsection{Phase Diagram}

From the thermal anomalies observed in the wide range of experimental data presented, a phase diagram can be sketched. As previously emphasized, all anomalies are broad, indicative of cross-overs between different regimes as opposed to sharp symmetry breaking phase transitions. It also means that different measurements will display slightly different characteristic temperatures. In combination though such measurements lead us to the definition of well defined cross over lines that we have labeled $T_{N}$, $T_{2}$, and $T_{3}$ bordering the corresponding phases I-III. The upper cross-over lines are remarkably field independent. The predominant effect of field appears to be broadening that can be associated with a spherically averaged result on an anisotropic material. At low temperatures a distinct non-Fermi liquid (NFL) regime is present for  H $\geq$ 2.5 T where $\rho(T)~\propto~\log T$ and $C_p~\propto~T^2$.

\begin{figure}[t!]
\includegraphics[width=1.0\columnwidth,clip]{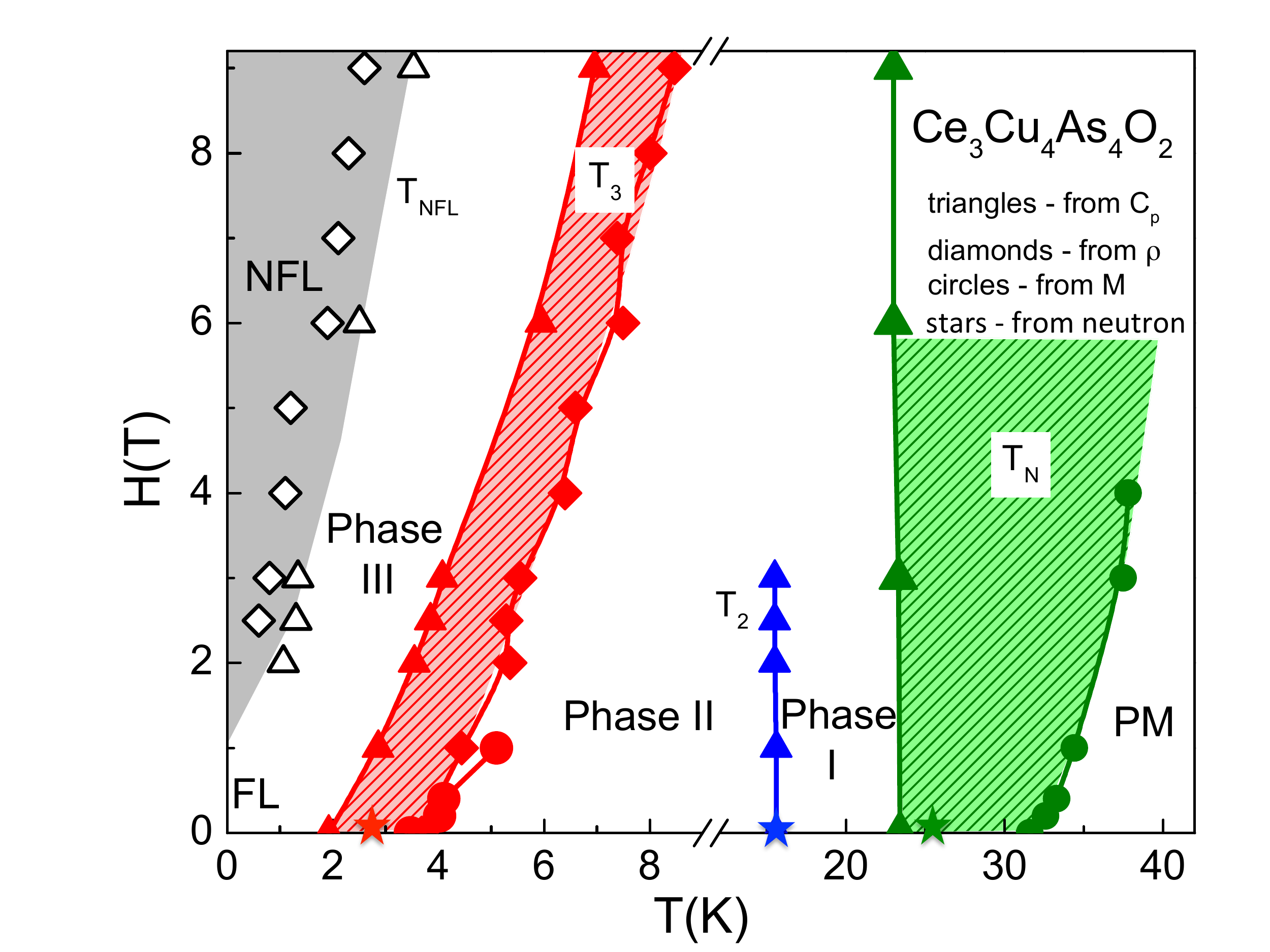}
\caption{\label{phase_diagram} Phase diagram of temperature vs. magnetic field for Ce$_3$Cu$_4$As$_4$O$_2$, with phase boundaries determined by features in the physical property measurements and neutron diffractions.  }
\end{figure}

\subsection{Magnetic Structures}
From the neutron scattering data presented in section \ref{mag_neutron_results} we have inferred the following principal features of magnetism in $\rm Ce_3Cu_4As_4O_2$: For $T< T_{N}=24$~K Ce spins are polarized along the tetragonal $\bf c$ axis, with a characteristic propagation vector $\textbf{k}_{1} = 001$. This structure persists in the second phase for $T~<~T_{2}~=~16$ K, which is marked by the loss of low energy spin fluctuations that have quasi-two-dimensional spatial correlations, a characteristic wave vector ${\bf k}_2=(0.5,0,0)$, and a component of polarization perpendicular to ${\bf k}_2$. Finally, for $T~<~T_{3}~=~2~K$ there is a spin-flop transition where the orientation of spins modulated in accordance with ${\bf k}_1$ rotates from the $\bf c$ axis to the tetragonal basal plane.

While the the present powder diffraction data are insufficient to definitely determine the complex spin structures of $\rm Ce_3Cu_4As_4O_2$, we shall develop a specific picture of the ordered spin structures that is consistent with the diffraction profiles and provides a logical sequence of phase transitions. One of the main outcomes that is supported by Rietveld analysis of the data is an association of the two wave vectors with the two different Ce sites in $\rm Ce_3Cu_4As_4O_2$. The  sequence of spin structures inferred from this analysis is shown in Fig.~\ref{structure}.

\subsubsection{Phase I}

We use representation theory to classify the magnetic structures with wave vector ${\bf k}_1=(001)$ that can develop from the paramagnetic phase of space group I4/mmm in a second order phase transitions. Using Kovalev notation \cite{Kova}, the representations associated with  ${\bf k}_1$ and each of the two cerium sites  decompose into irreducible representations (IR) as follows. For the Ce1 cite associated with  $\rm CeRu_4As_4$ layers: $\Gamma_{mag}(\rm Ce1) = \Gamma_3 + 2\Gamma_9$. For the Ce2 sites associated with the $\rm Ce_2O_2$ layers: $\Gamma_{mag}(\rm Ce2) = \Gamma_3 + \Gamma_2 + 2\Gamma_9 + 2\Gamma_{10}$.

There are a total of nine basis vectors (BVs) and these are listed in Table \ref{BVs}. These were obtained using the \textit{SARAh} program \cite{sarah}. For both cerium sites the two-dimensional $\Gamma_9$  and $\Gamma_{10}$  IRs describe magnetic structures with in-plane moments while the one-dimensional $\Gamma_3$   and $\Gamma_2$ IRs describe structures with moments along the tetragonal c-axis.

\begin{table}
\setlength{\tabcolsep}{8pt}
\caption{The nine basis vectors associated with magnetic structures that transform according to irreducible representations of the  I \textit{/4mmm} space group with propagation vector \textbf{k}=(0,0,1). The irreducible representations and basis vectors were obtained using \textit{SARAh} \cite{sarah} and the notation is based on Kovalev\cite{Kova}. }
\begin{tabular}{| c | c | c | c | c |}
\hline
\multirow{2}{*}{IR} & \multirow{2}{*}{Atom}     & \multicolumn{3}{l|}{Basis Vectors} \\ \cline{3-5}
                                                     &                                           & $m_a$        & $m_b$      & $m_c$     \\ \hline
$\Gamma_3$                                &      $\rm Ce_1$                         &    0    &     0    &     1  \\
                                                     &     $\rm Ce_2$                          &    0    &     0    &     1  \\ \hline
\multirow{4}{*}{$\Gamma_9$}   &  \multirow{2}{*}{$\rm Ce_1$} &   1     &     0    &     0  \\ \cline{3-5}
                                                     &                                            &   0     &    -1    &     0  \\ \cline{2-5}
                                                     & \multirow{2}{*}{$\rm Ce_2$}  &   1     &     0    &     0  \\ \cline{3-5}
                                                     &                                            &    0    &    -1    &     0   \\ \hline
$\Gamma_2$                                &         $\rm Ce_2$                        &    0    &     0    &    -1   \\ \hline
\multirow{2}{*}{$\Gamma_{10}$ } & \multirow{2}{*}{ $\rm Ce_2$} &   -1    &     0    &    0     \\ \cline{3-5}
                                                     &                                             &    0    &     1    &    0      \\ \hline
\end{tabular}

\label{BVs}
\end{table}

Because no magnetic peaks of the form $(0,0,2n+1)$ are observed for $T_{2}<T<T_{N}$ (Fig.~\ref{diffraction_pattern}), the moment is oriented along the c-axis and we must focus on the one-dimensional IRs $\Gamma_2$ and $\Gamma_3$. For the Ce1 sites this leaves just one option namely $\Gamma_3$, which corresponds to the centered site antiparallel to the nearest neighbors (Fig.~\ref{structure}). For Ce2 sites there are two options corresponding to FM ($\Gamma_2$) or AFM ($\Gamma_3$) alignment of spins within a $\rm Ce_2O_2$ bi-layer.

Assuming the $\Gamma_2$ representation for Ce2 sites only (and no order on Ce1 sites) leads to $\chi^2=23.7$, which is larger than $\chi^2=17.5$ corresponding to $\Gamma_3$ with order on both cerium sites. Here the reduced $\chi^2$ goodness is reported for difference data in the range $1.5\AA^{-1}<Q<1.8\AA^{-1}$.  Fig.~\ref{detailph} shows three fit lines corresponding to $\Gamma_3$ order on Ce2, on Ce1 sites only and on both sites.The constrained differential line shape associated with thermal expansion is included with $\Delta a/a=1.3\%$ and $\Delta c/c=0.6\%$. The fit with order on both sites provides the best account for the relative intensity of the two peaks and the corresponding ordered moments inferred are $\mu_{Ce_1} =0.14(6)~\mu_B $ and $\mu_{Ce_2} = 0.18(2)~\mu_B$. A sketch of the corresponding spin configuration is shown as phase I in Fig.~\ref{structure} (b). The weak nuclear Bragg peak (101) is close to the magnetic peaks. Subtle changes in the chemical structure thus could also play a role in the temperature dependence of diffraction in this Q-range.

\subsubsection{Phase II}

While the present experiment detects no new elastic peaks upon entering phase II, a distinct loss of low energy inelastic scattering is observed in the energy integrating detectors of the MACS instrument for $T<T_{2}$ (Fig.~\ref{Neutron2}(b-c)). In Fig.~\ref{Neutron3} the broad anisotropic peaks associated with the scattering that vanishes for $T<T_{\rm 2}$ is compared to the spherical average of the following dynamic correlation function associated with quasi-two-dimensional magnetic correlations:
\begin{equation}
{\cal S}(\textbf{Q})= \sum_m \frac{\langle S^2\rangle \cdot \xi^2/\pi}{[1+(\xi|{\bf Q_{\perp}-Q_m}|)^2]^2}.
\end{equation}
Here ${\bf Q}_m = \tau \pm \textbf {k}_{2}$ indicates the location of critical points in the 2D reciprocal lattice. Because the incident neutron energy is similar to the energy scale of the spin system,  actual wave vector transfer differs from that calculated neglecting energy transfer as follows
\begin{equation}
\tilde{Q}(\theta,\omega) =   k_{i}\sqrt{2-\frac{\hbar\omega}{E_i}-2\sqrt{1-\frac{\hbar\omega}{E_i}}\cos 2\theta}.
 \end{equation}
With $\hbar\omega=1.2(3)$~meV, in-plane correlations length $\xi = 8.2(6)  \AA$ and an effective fluctuating moment of $(g_J\mu_B)^2\langle S^2\rangle\leq0.3~\mu_B^2$ this model provides a satisfactory account of the difference data in Fig.~\ref{Neutron3}. Here we have plotted the data versus the inferred wave vector transfer for inelastic scattering $\tilde{Q}$ showing also wave vector transfer for elastic scattering on the upper horizontal axis.

Ce2 sites play the dominant role both in Phase I and Phase III. We therefore pursue the hypothesis that the ${\bf k}_2=(1/2,0,0)$ type scattering in Fig.~\ref{Neutron3} is associated with Ce1 sites. The total scattering can decrease if the kinematically accessible Bragg peaks of the corresponding order are extinguished by the polarization factor. Fig.~\ref{diffraction_pattern} (b) includes the calculated elastic scattering that would result from the longitudinally polarized in-plane antiferromagnetic spin structure indicated in Fig.~\ref{structure}(b) for Phase II. Because spins are almost parallel to ${\bf k}_2$ at low wave vector transfer, this structure produces very little magnetic powder diffraction and would go undetected in the present experiment. This structure is therefore a viable candidate to account for the loss of critical scattering with wave vector ${\bf k}_2$ without the appearance of appreciable magnetic Bragg scattering.

\subsubsection{Phase III}

The appearance of low-$Q$ peaks in Phase III of the ${\bf k}_1=(001)$ variety is indicative of in-plane magnetic moments that are described by IR $\Gamma_9$ or $\Gamma_{10}$ (Table~\ref{BVs}). Of these $\Gamma_{10}$ is inconsistent with the observed relative intensities while $\Gamma_9$ provides for the excellent fit shown in Fig.~\ref{diffraction_pattern}(c). The contribution to magnetic scattering from out of plane moments (Fig.~\ref{diffraction_pattern}(a-b)) is so weak that we cannot place meaningful limits on how much might remain in Phase III. The experimental data in Fig.~\ref{diffraction_pattern}(c) is consistent with an AFM spin structure of the $\Gamma_9$ variety (Fig.~\ref{structure}(b)) with  moments within the tetragonal basal plane. The total moment obtained is ($\mu_{Ce_1}+\mu_{Ce_2}) =0.85(2) \mu_B$. Consistent with the assignment of Ce1 with moment  $\leq0.45 \mu_B$ to the ${\bf k}_2$ structure, the Rietveld refinement displayed in Fig.~\ref{diffraction_pattern}(c) indicates the Ce1 site contributes less than 0.1 $\mu_B$ to the ${\bf k}_1$ structure of Phase III.

In a field of 7 Tesla the magnetization per cerium site reaches a value of 0.6~$\mu_B$ corresponding to 2/3 of the low $T$ ordered moment on the Ce2 sites, which make up 2/3 of the cerium in $\rm Ce_3Cu_4As_4O_2$. This is consistent with the magnetization in 7 T of only the Ce2 sites that consist of FM layers leaving the Ce1 layers in the proposed in-plane ${\bf k}_2=(0.5,0,0)$ type AFM order. This hypothesis could be examined by higher field magnetization measurements.

\subsection{Electronic Structure}

Density functional calculations were done on Ce$_3$Cu$_4$As$_4$O$_2$ for three types of magnetic configurations. The first is the ferromagnetic (FM) state. The second is the AFM (0, 0, 1) periodicity, which corresponds to phase I in Fig. \ref{structure} and is referred to as AFM1. The third magnetic configuration is to model phase III in Fig. \ref{structure}, with the same magnetic moment arrangements for Ce2 sites as in the AFM1 state and (1/2, 0, 0) periodicity for Ce1 sites. This structure is referred to as AFM2. The magnetic unit cell for the AFM2 state is $2\times1\times1$ times the crystallographic unit cell.

Among all three magnetic configuration considered, AFM2 has the lowest energy. The relative energy for AFM2 state is E(AFM2) - E(FM) = - 127 meV/F.U., while the relative energy for AFM1 is E(AFM1) - E(FM) = 2 meV/F.U. The energy comparison directly supports AFM2 as the ground state, which is consistent with the neutron scattering result. For the AFM2 state, the magnetic moments on both Ce sites are between 0.96 $\sim$ 0.97 $\mu_B$, significantly larger than the experimentally observed values on Ce1 sites (0.45 $\mu_B$), but close to Ce2 sites (0.85 $\mu_B$). One possible reason for the reduced moment on Ce1 compared to DFT is that the Ce1 atoms and As2 atoms are both spatially and energetically close to each other, resulting in Kondo hybridization between them. Such Kondo hybridization can strongly reduce the sizes of magnetic moments \cite{weibao, kawarazaki}.

\begin{figure}[t!]
\includegraphics[width=1.0\columnwidth,clip]{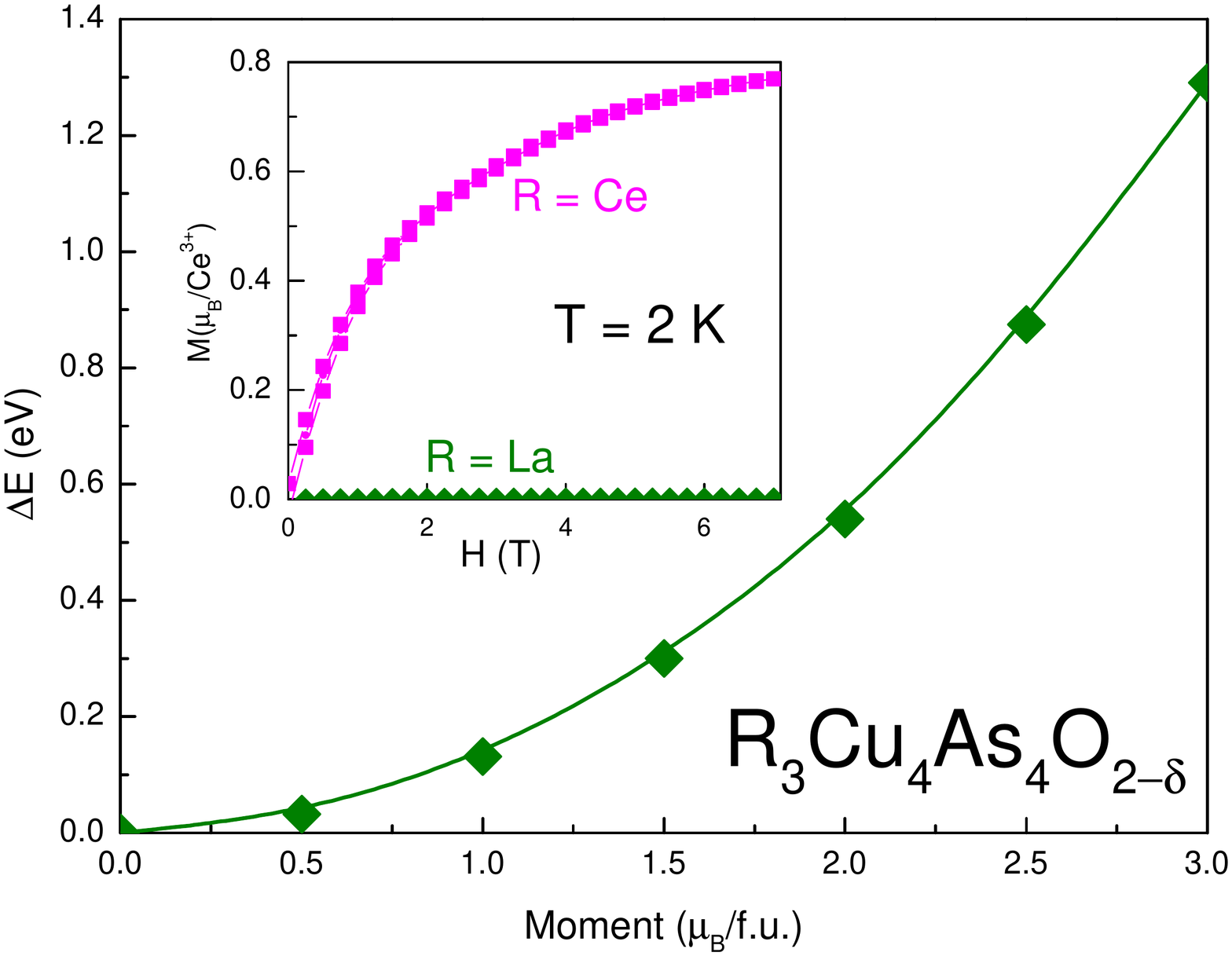}
\caption{\label{fixed_moment} Energy versus magnetic moment from the fixed spin moment calculation for La$_3$Cu$_4$As$_4$O$_2$, the non-magnetic analogue of Ce$_3$Cu$_4$As$_4$O$_2$. Inset: M(H) isotherm for both Ce$_3$Cu$_4$As$_4$O$_2$ and La$_3$Cu$_4$As$_4$O$_2$.  }
\end{figure}

\begin{figure}[t!]
\includegraphics[width=1.0\columnwidth,clip]{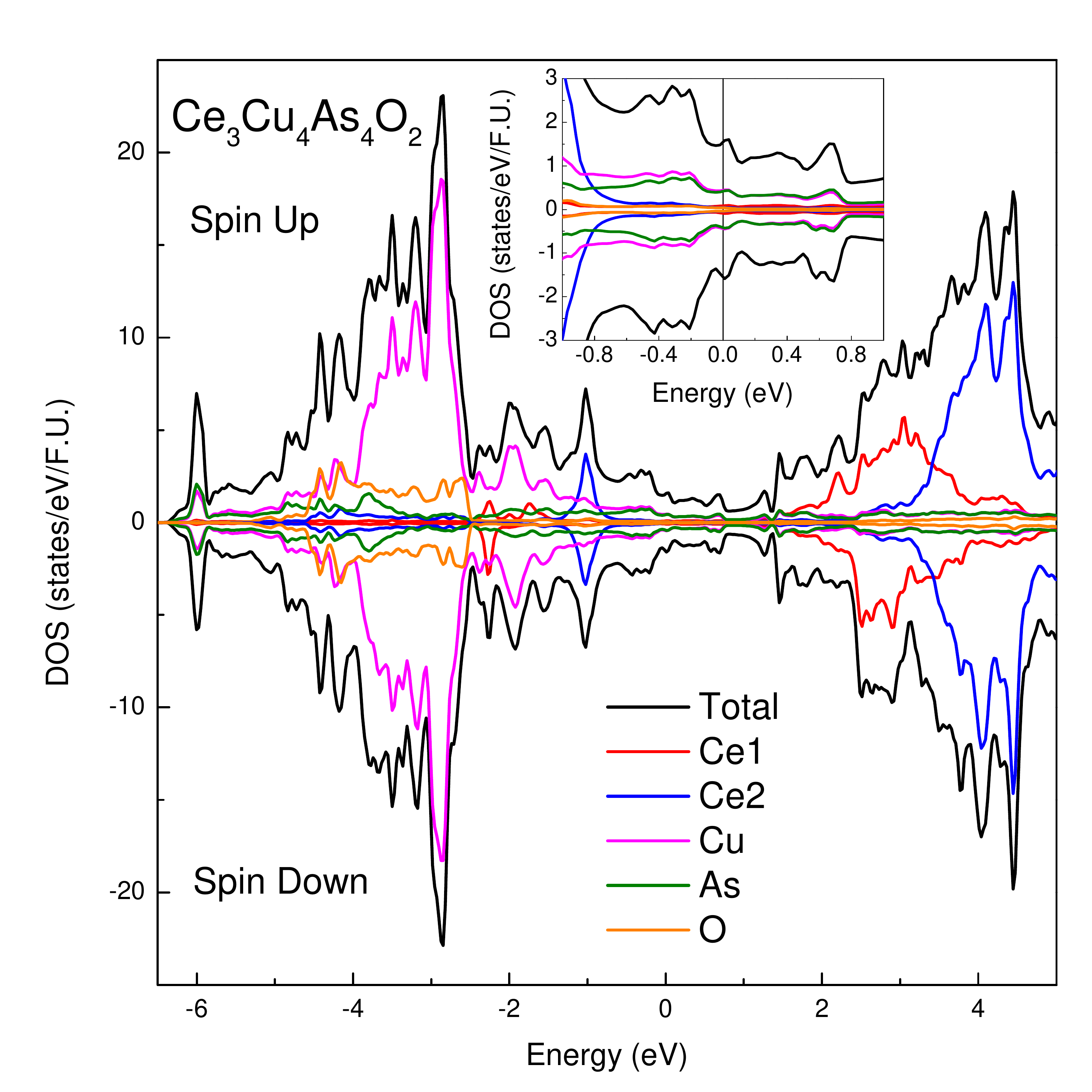}
\caption{\label{DOS} The total and atom-projected DOS calculated for Ce$_3$Cu$_4$As$_4$O$_2$ in the AFM2 state. The spin down DOS is multiplied by -1. Inset is an enlarged view for DOS close to the Fermi level. }
\end{figure}

\begin{figure}[t!]
\includegraphics[width=0.8\columnwidth,clip]{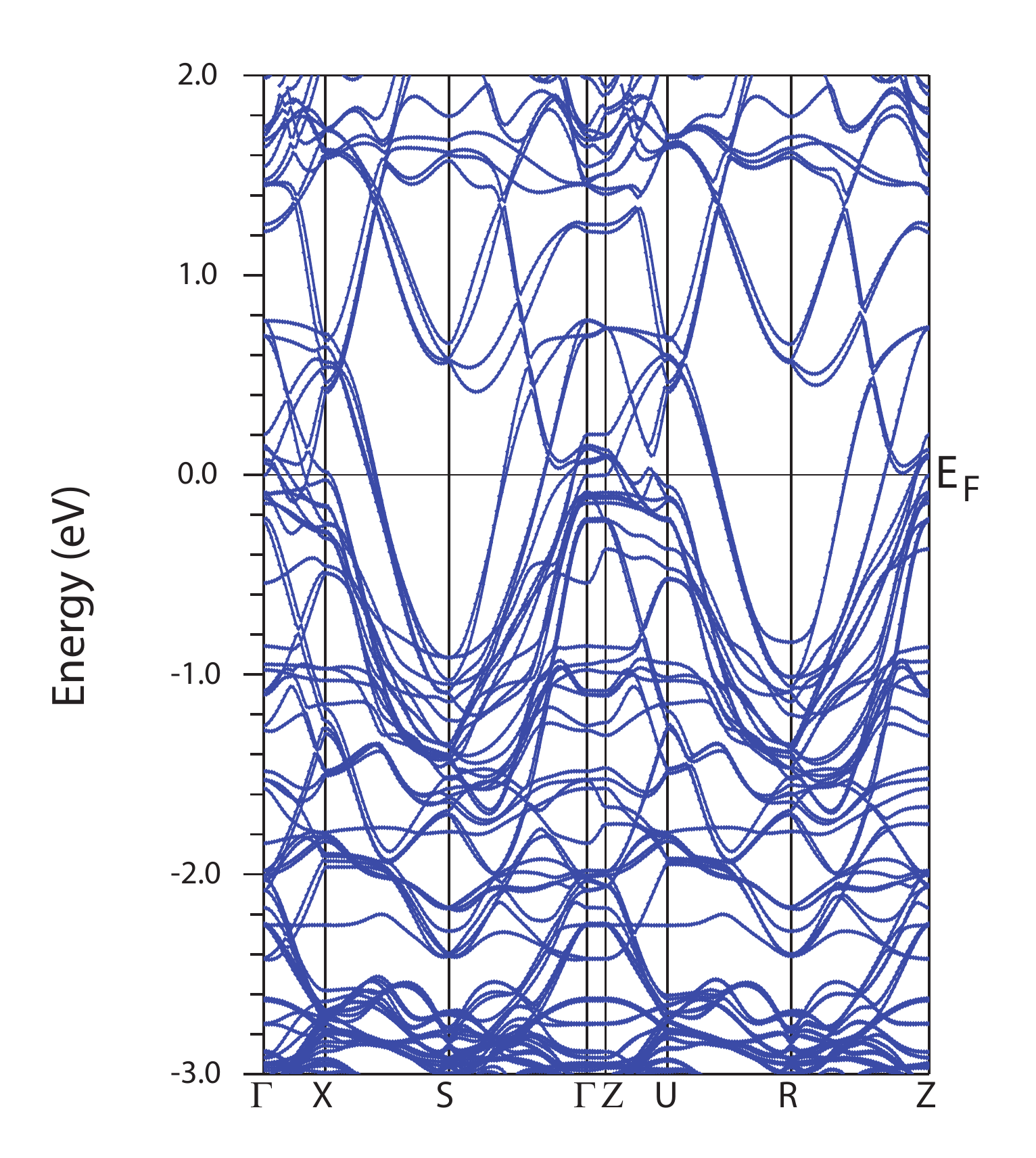}
\caption{\label{band_structure} The band structure calculated along an in-plane k-path for Ce$_3$Cu$_4$As$_4$O$_2$ in the AFM2 state, with only the spin up component shown.}
\end{figure}

\begin{figure}[t!]
\includegraphics[width=1.0\columnwidth,clip]{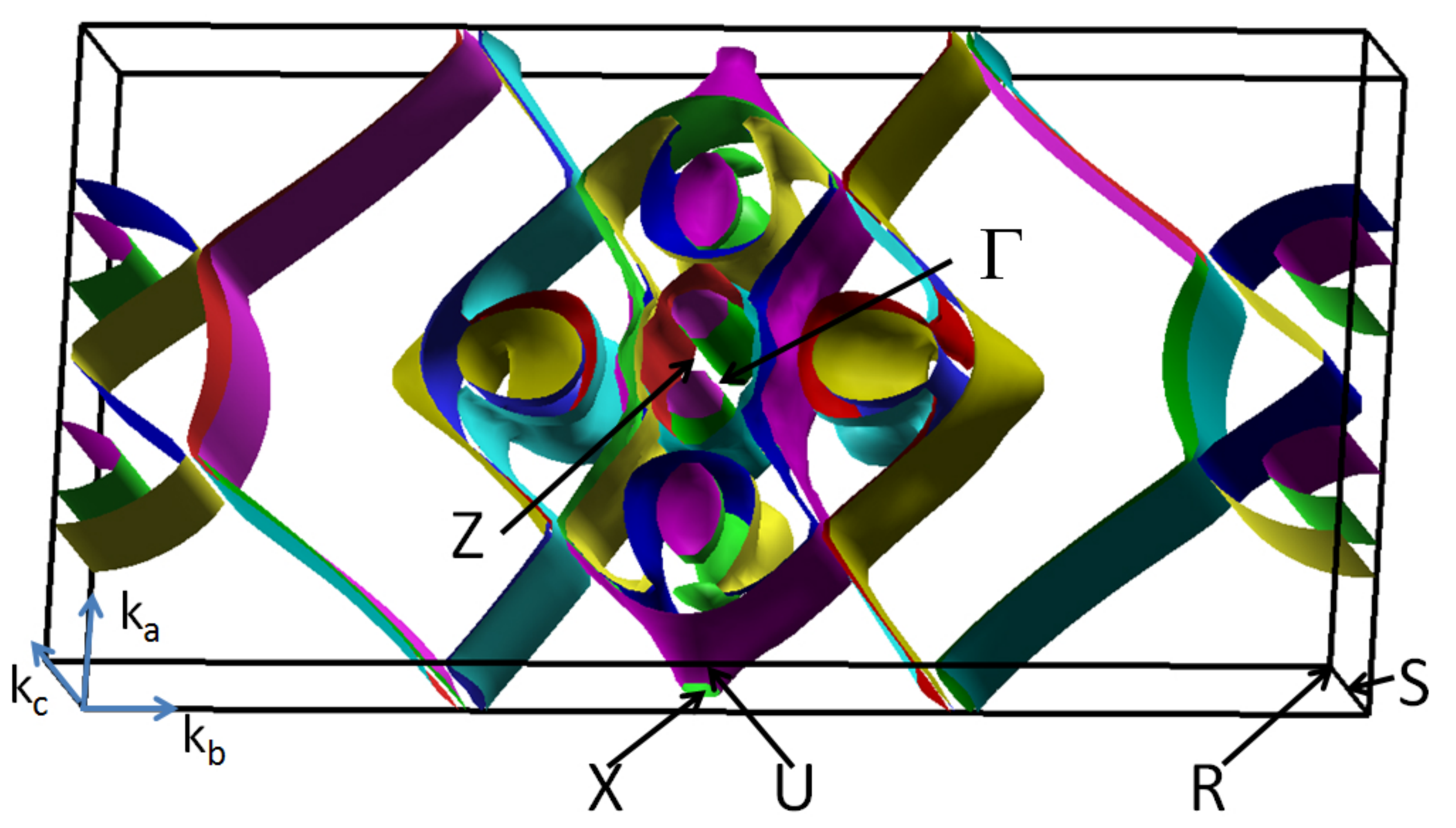}
\caption{\label{fermi_surface} The Fermi surface for Ce$_3$Cu$_4$As$_4$O$_2$ in the AFM2 state(spin up component only).}
\end{figure}

To examine the magnetism on the Cu site, a fixed spin moment calculation was performed for La$_3$Cu$_4$As$_4$O$_2$ (Fig. \ref{fixed_moment}), the non-magnetic analogue of Ce$_3$Cu$_4$As$_4$O$_2$, with methods similar to those for the Ce compound. The energy of La$_3$Cu$_4$As$_4$O$_2$ increases monotonically with increasing spin moment, indicative of a paramagnetic ground state. This monotonic behavior implies that there is no tendency for the 3d electrons to become magnetic, which is consistent with the experimental observations\cite{jiakui}.

The density of states (DOS) for Ce$_3$Cu$_4$As$_4$O$_2$ in the AFM2 state is shown in Fig. \ref{DOS}. Unlike for Fe-pnictides case, the majority of the Cu bands lie in the energy interval close to the O bands, despite the fact that O has lower electronegativity than Cu. The increased energy of O bands might be due to their proximity to Ce atoms. At the Fermi level, Cu and As contribute most to the DOS. The non-zero DOS at the DFT Fermi level is indicative of a metal. The Ce1 atoms have a lower energy, and their bands are broader than for the Ce2 atoms.

The Cu 3d band is very broad, with a bandwidth of about 10 eV. Although centered around -3 eV, the tail of the 3d band is visible up to 5 eV, and the Fermi level just crosses this tail. This indicates a weakly correlated 3d band, consistent with a non-magnetic state for Cu.

The band structure for Ce$_3$Cu$_4$As$_4$O$_2$ in the AFM2 state is shown in Fig. \ref{band_structure}. A quasi-two-dimensional nature is apparent from the similarity of the bands along the in-plane $\Gamma$-X-S-$\Gamma$ path and Z-U-R-Z paths. As for the iron pnictides, most bands crossing the Fermi levels are anti-bonding \cite{singh} and associated with Cu or As. The flat band near -2.1 eV is from Ce1, while the less visible Ce2 band is centered around -1.0 eV.

The Fermi surface is shown in Fig. \ref{fermi_surface} plotted in the folded Brillouin zone corresponding to the orthorhombic magnetic unit cell which is $2\times1\times1$ times the tetragonal crystallographic unit cell. Extended along the c-axis with almost no corrugation, the Fermi surface indicates electron hopping is mainly confined to the tetragonal basal plane so that that the electronic state for Ce$_3$Cu$_4$As$_4$O$_2$ is very two dimensional. The pockets near the $\Gamma$ points are however more dispersive along c-axis than the pockets far away the $\Gamma$ points.

\begin{table*}
\setlength{\tabcolsep}{6pt}
\caption{Our working hypothesis for the magnetic structure in the three distinct phases of $\rm Ce_3Cu_4As_4O_2$: phase \textit{I} (  $16~{\rm K}  < T <24$~K), and phase \textit{II} ( $ 2~{\rm K} < T < 16$~K ) and phase \textit{III} ($ T < 2$~K  ). In the table, $\bf k_1$ = (0, 0, 1) and $\bf k_2$ = (1/2, 0, 0). One dimensional IR $\Gamma_2$ associated with $\bf k_2$ represents  Basis Vectors as (100).  The hypothesis is consistent with the experimental data but more specific than can be proven on their basis.  }
\label{summary}

  \begin{tabular}{|l|l|c|c|c|c|c| }
  \hline
    \multirow{2}{*}{  \textbf{Site}	} & 	\multirow{2}{*}{  \textbf{Property} }	& \multicolumn{2}{c|}{  \textbf{ Phase \textit{III} } }     &  \multicolumn{2}{c|}{ \textbf{ Phase \textit{II} } } &  \textbf{ Phase \textit{I} }
\\  & & \multicolumn{2}{c|}{ $ T < 2$~K  } &  \multicolumn{2}{c|}{ $ 2~{\rm K} < T < 16$~K } &  $16~{\rm K}  < T <24$~K
      \\\hline						

   \multirow{4}{*}{$\rm Ce1$  in $\rm CeCu_4As_4$ }   	&		Propagation vector ($\bf k$) 	&$\bf k_1$         &  $\bf k_2$  	& $\bf k_1$			& $\bf k_2$  			  &  $\bf k_1$ 	  \\
									&        Classification (IR)				&  $\Gamma_9$ &  $\Gamma_2$  &	$\Gamma_3$  	& 	$\Gamma_2$  	& $\Gamma_3$   	 \\
                                              					&	Spin orientation				&  \textbf{b}  	&	 \textbf{a}		& 	\textbf{c} 		&  \textbf{a} 	&\textbf{c} 	           \\
									&	Moment size ($\mu_B$)			& $< 0.1$  	& $\leq0.45$			      &    0.17(2)			& $\leq$0.3 	 & 	0.14(6)	\\ \hline

  \multirow{4}{*}{$\rm Ce2$ in $\rm Ce_2O_2$  }   		&		Propagation vector ($\bf k$) 	& \multicolumn{2}{c|}{$\bf k_1$}      	&  \multicolumn{2}{c|}{$\bf k_1$}		 & {$\bf k_1$ }	  \\
									&      Classification (IR)				&\multicolumn{2}{c|}{ $\Gamma_9$ }     & \multicolumn{2}{c|}{ $\Gamma_3$ 	}        &{$\Gamma_3$} 	    \\
									&	Spin orientation				& \multicolumn{2}{c|}{ \textbf{b}}	 &  \multicolumn{2}{c|}{ \textbf{c}}  		&{  \textbf{c}} 		\\
									& 	Moment size ($\mu_B$)  			&\multicolumn{2}{c|}{0.85(1)}		 &\multicolumn{2}{c|}{0.22(3)}			&{0.18(2)}   			 \\ \hline

 \end{tabular}

\end{table*}

\section{Discussion}

\label{discussion}

Having presented and analyzed comprehensive data for the magnetic and electronic properties of $\rm Cu_3Cu_4As_4O_2$ we now discuss the overall physical picture of this material. Table~\ref{summary} provides a summary of the spin structures proposed for the three magnetic phases. While the upper two transitions mark the development of distinct magnetic modulations with wave vectors ${\bf k}_1$ = (001) and ${\bf k}_2$ = (1/2, 0, 0) in the $\rm Ce_2O_2$ and the $\rm CeCu_4As_4$ layers respectively, the transition to phase III is a spin flop transition where the staggered magnetization already established for $\rm Ce_2O_2$ in phase I rotates into the basal plane. This last transition is by far the most prominent as observed by magnetic neutron scattering because it allows for diffraction at low $Q$ which is extinguished by the polarization factor in the higher temperature phases.

The lower transition to phase III might be understood as follows. If we allow for isotropic bilinear spin exchange interactions between different cerium sites then the effect of the $\rm Ce_2O_2$ layers on the $\rm CeCu_4As_4$ layers is analogous to a uniform magnetic field, which in phase II is oriented perpendicular to the staggered magnetization and perpendicular to the basal plane. Rotation of the $\rm Ce_2O_2$ magnetization into the basal plane can then be understood as the result of a competition between incompatible single ion and/or exchange anisotropy for the Ce1 and Ce2 sites. Having ordered antiferromagnetically with spins within the basal plane in phase II, $\rm CeCu_4As_4$ layers apparently have an easy plane character. Phase I on the other hand demonstrates an easy axis character to $\rm Ce_2O_2$ layers. The  final spin flop transition might then be understood as the easy-plane character prevailing if $\rm Ce_2O_2$ spins rotate into the basal plane and perpendicular to ${\bf k}_2$. This is consistent with the diffraction data though the diffraction data do not establish the orientation of $\rm Ce_2O_2$ spins within the basal plane.

Comparison to $\rm Pr_3Cu_4As_4O_{2-\delta}$ is instructive. The strength of RKKY exchange interactions can be expected to vary between rare earth ions in accordance with the so-called de Gennes factor $F=(g-1)^2J(J+1))$, which is 0.18 for cerium and 0.8 for praseodymium for a ratio of $F({\rm Pr})/F({\rm Ce})=4.4$. While the ordered spin structures for $\rm Pr_3Cu_4As_4O_{2-\delta}$ are as yet unknown, there are two phase transitions at $T_{N}=35(3)$~K and $T_{2}=22$~K \cite{jiakui}. The enhancement of $T_N$ is not nearly as large as might be expected based on de Gennes scaling but the similar ratio for $T_{2}/T_{N}$ of approximately 1.5 for Ce and 1.6 for Pr is consistent with an analogy between phases I and II of the two compounds. The absence of a third transition for the praseodymium compound is consistent with $T_{3}$ for $\rm Ce_3Cu_4As_4O_{2}$  being contingent on competing anisotropies that might be absent for $\rm Pr_3Cu_4As_4O_{2-\delta}$.

Replacing Ni for Cu in $\rm CeNi_4As_4O_{2-\delta}$ results in just a single peak in the specific heat at $T_{N}=1.7$~K\cite{jiakui}. The change in entropy up to 40~K, the effective moment, and the Weiss temperature are however, similar to $\rm CeCu_4As_4O_2$, which indicates the cerium moment persists with similar interaction strengths. Increased magnetic frustration and quenched chemical or structural disorder are potentially relevant factors to account for the suppression relative to  $\rm Ce_3Cu_4As_4O_{2}$ of the higher temperature transitions. Here we note that the rare earth layers are separated by transition metal ions so that all inter-layer interactions must proceed through the copper or nickel layer. The dramatic effect of changing these layers from copper to nickel thus might indicate  the upper transitions are driven by interlayer interactions.

A curious feature of $\rm Ce_3Cu_4As_4O_{2}$ are the broad nature of the specific anomalies. The diffraction experiments presented here reveal magnetic order above the 75 $\AA$ length scale but in harmony with the specific heat data, the corresponding staggered magnetization emerges gradually upon cooling and without a  critical onset (Fig.~\ref{Neutron2} and Fig.~\ref{Neutron4}). These specific heat and scattering data  consistently point to cross-over phenomena rather than actual phase transitions and thus an absence of true symmetry breaking. Potentially relevant features that could lead to this unusual situation are (1) disorder, particularly for the weak interactions between layers (2) frustration (3) the interplay of alternating layers of spins with distinct quasi-2D Ising transitions and (4) Kondo screening. To distinguish between these scenarios and achieve a deeper understanding of the unusual sequence of transitions in $\rm Ce_3Cu_4As_4O_{2}$ will require single crystalline samples from which more specific information can be obtained via magnetic neutron diffraction.

\section{Conclusions}
\label{conclusion}

In conclusion, field dependent physical properties and neutron scattering measurements as well as band structure calculations were carried out on the layered transition metal pnictide compound $\rm Ce_3Cu_4As_4O_{2}$. All investigations point to a quasi-two-dimensional electronic state and modulated complex magnetic structure. The crystal structure resembles an interleaving 122 and 1111 Fe pnictide structures, resulting in two rare earth sites with very different local environments, which in turn lead to the complex magnetic structure. The neutron scattering measurements reveal three successive transitions, consistent with the magnetization and specific heat measurements. For the first transition at T$_{N}$ = 24 K, neutron scattering indicates alternative FM layers with spins oriented along c. Below the second transition T$_{2}$ = 16 K neutron scattering reveals the loss of spin fluctuations which we can account for by in-plane AFM ordering of Ce1 site. The third transition $T_{3}$ = 1.9 K appears to be a spin-flop transition where all the magnetic moment directions switch to in-plane polarization. Band structure calculation confirm the AFM2 state, which models the magnetic phase below $T_{3}$. A notably large negative magneto-resistance ($\approx$-30 $\%$) around $T_{3}$ is observed up to H = 9 T, implying significant impact for magnetic fluctuations on the transport behavior. Under high magnetic field H $>$ 3 T, the temperature dependence for both resistivity and heat capacity violate Fermi liquid behavior. This is possibly due to the anisotropic quasi-two-dimensional nature of the electronic state as indicated by neutron data and the band structure calculations.

\section{Acknowledgements}
The work at Rice University was supported by AFOSR MURI. Work at IQM was supported by the US Department of Energy, office of Basic Energy Sciences, Division of Material Sciences and Engineering under grant DE-FG02-08ER46544. This work utilized facilities supported in part by the National Science Foundation under Agreement No. DMR-0944772. The authors thank Andriy Nevidomskyy, Meigan Aronson and Liang Zhao for useful discussions.


\begin{thebibliography}{99}
\bibitem{quantumdesign} The identification of  the commercial product does not indicate that it is endorsed by NIST or is recommended by NIST or that it is necessarily the best equipment for the purposes  described.
\end{thebibliography}
\end{document}